\documentclass[preprint,12pt]{elsarticle}

\makeatletter
\def\ps@pprintTitle{%
  \let\@oddhead\@empty
  \let\@evenhead\@empty
  \def\@oddfoot{\footnotesize\itshape
       \hfill\today}%
  \let\@evenfoot\@oddfoot}
\makeatother
\usepackage{amssymb}
\usepackage{amsmath}
\usepackage{hyperref}
\usepackage{xcolor}

\usepackage{subcaption}

\usepackage[nolist,nohyperlinks]{acronym}
\begin{acronym}[MPTPS]
  \acro{EISA}{evaporation-induced self-assembly}
  \acro{NIPS}{nonsolvent-induced phase separation}
  \acro{ML}{Machine-learning}
  \acro{MLP}{multilayer perceptron}
  \acro{SNIPS}{solvent evaporation and nonsolvent-induced phase separation}
  \acro{CUDA}{compute unified device architecture by Nvidia\textregistered}
  \acro{FFT}{fast Fourier transformation}
  \acro{GP}{Gaussian processes}
  \acro{GPU}{graphics processing unit}
  \acro{MCS}{Monte Carlo sweep}
  \acro{MPI}{message passing interface}
  \acro{SCMF}{single-chain-in-mean-field}
  \acro{SOMA}{SOft coarse-grained Monte carlo Acceleration}
  \acro{TPS}{Time steps Per Second}
  \acro{UDM}{Uneyama-Doi model}
  \acro{WSL}{weak-segregation limit}
  \acro{SI}{supplementary information}
  \acro{SCFT}{self-consistent field theory}
\end{acronym}

\newcommand{\Nbar}{\ensuremath{\bar{\mathcal N}}}
\newcommand{\kT}{k_{\textrm B}T}

\begin{document}

\begin{frontmatter}

%% Title, authors and addresses

%% use the tnoteref command within \title for footnotes;
%% use the tnotetext command for theassociated footnote;
%% use the fnref command within \author or \affiliation for footnotes;
%% use the fntext command for theassociated footnote;
%% use the corref command within \author for corresponding author footnotes;
%% use the cortext command for theassociated footnote;
%% use the ead command for the email address,
%% and the form \ead[url] for the home page:
%% \title{Title\tnoteref{label1}}
%% \tnotetext[label1]{}
%% \author{Name\corref{cor1}\fnref{label2}}
%% \ead{email address}
%% \ead[url]{home page}
%% \fntext[label2]{}
%% \cortext[cor1]{}
%% \affiliation{organization={},
%%             addressline={},
%%             city={},
%%             postcode={},
%%             state={},
%%             country={}}
%% \fntext[label3]{}

\title{Machine-learned domain partitioning for computationally efficient coupling of continuum and particle simulations of membrane fabrication}
%\title{Machine learned error prediction based domain partitioning for adaptive multi fidelity simulation coupling}% working title, open for suggestions

\author[1]{Matthias Busch\corref{cor}}\ead{matthias.busch@tuhh.de}
\author[2]{Gregor Häfner\corref{cor}}\ead{gregor.haefner@uni-goettingen.de}
\author[2]{Jiayu Xie}
\author[3]{Marius Tacke}
\author[2]{Marcus Müller}
\author[1,3]{Christian J. Cyron}
\author[1,3]{Roland C. Aydin\corref{cor}}\ead{roland.aydin@hereon.de}

\affiliation[1]{organization={Institute for Continuum and Material Mechanics, Technical University of Hamburg}, addressline={Eißendorfer Straße 42}, city={Hamburg}, postcode={21073}, country={Germany}}

\affiliation[2]{organization={Institute for Theoretical Physics, Georg August University Göttingen}, addressline={Friedrich-Hund-Platz 1}, city={Göttingen}, postcode={37077}, country={Germany}}

\affiliation[3]{organization={Institute of Material Systems Modeling, Helmholtz-Zentrum Hereon}, addressline={Max-Planck-Straße}, city={Geesthacht}, postcode={21502}, country={Germany}}

\cortext[cor]{Corresponding authors}

%% Abstract
\begin{abstract}
All simulation approaches eventually face limits in computational scalability when applied to large spatiotemporal domains. This challenge becomes especially apparent in molecular-level particle simulations, where high spatial and temporal resolution leads to rapidly increasing computational demands. To overcome these limitations, hybrid methods that combine simulations with different levels of resolution offer a promising solution. In this context, we present a machine learning–based decision model that dynamically selects between simulation methods at runtime. The model is built around a \ac{MLP} that predicts the expected discrepancy between particle and continuum simulation results, enabling the localized use of high-fidelity particle simulations only where they are expected to add value. This concurrent approach is applied to the simulation of membrane fabrication processes, where a particle simulation is coupled with a continuum model. This article describes the architecture of the decision model and its integration into the simulation workflow, enabling efficient, scalable, and adaptive multiscale simulations.
\end{abstract}

% %%Graphical abstract
% \begin{graphicalabstract}
% \centering
% \includegraphics[width=\linewidth]{graphics/GraphicalAbstractMLPaperV5.png}
% \end{graphicalabstract}

% %%Research highlights
% \begin{highlights}
% % 85: 123456789 123456789 123456789 123456789 123456789 123456789 123456789 123456789 12345
% \item Large time and length scales of membrane fabrication can be simulated with continuum and particle models
% \item By domain partitioning, both approaches can be coupled to minimize computational costs
% \item Machine learning enables error-based, dynamic domain partitioning
% \item Descriptor-based architectures support variable sizes of domains
% \item Iterative neural net application enhances generalizability over the time horizon
% \end{highlights}

%% Keywords
% \begin{keyword}
% Multi-fidelity simulation\sep Machine learning\sep Error prediction\sep SOMA\sep UDM
% %% keywords here, in the form: keyword \sep keyword

% %% PACS codes here, in the form: \PACS code \sep code

% %% MSC codes here, in the form: \MSC code \sep code
% %% or \MSC[2008] code \sep code (2000 is the default)

% \end{keyword}

\end{frontmatter}

%% Add \usepackage{lineno} before \begin{document} and uncomment 
%% following line to enable line numbers
%% \linenumbers

\section{Introduction}
% broad introduction with literature review to scale bridging, ml descriptors on simulation/error prediction and the simulations
% Explain goal of this paper: Show the decision model for this exemplary case such that others can use it for their own use case

%\MM{added Jaiyu Xie as co-author; added Gregor's email address}

Fabrication of diblock copolymer membranes is a challenging process \cite{Muller2021Nov}. Computer simulations are a valuable tool to optimize this process. Such simulations can rely on different approaches. On the one hand, phase-field continuum models can be used. They are computationally efficient, but lack resolution of details on the molecular scale. On the other hand, particle simulations are possible. They capture many more details but are also computationally much more expensive.  This limits their use in large‐scale applications such as membrane fabrication, biological assemblies, or complex fluids in realistic geometries. To address this challenge, one can use multiscale, multi-fidelity, and hybrid modeling techniques. Generally, these can couple fine‐grained, high-fidelity representations (like particle models) and coarse-grained, low-fidelity representations (like continuum models), activating the former only where necessary, thereby increasing  the global computational efficiency. Seminal methods include the heterogeneous multiscale method (HMM) \cite{E2003, Muller2011Nov}, the quasi‐continuum (QC) approach \cite{Tadmor1996}, and coupled atomistic/continuum (CAC) schemes \cite{Belytschko2003}.

However, traditional coupling schemes typically use heuristic triggers -- e.g., stress, strain or composition gradients -- to switch fidelity, which can overlook subtle error excursions or over-resolve low-error regions. \ac{ML} offers a data‐driven alternative, enabling surrogate models to predict the local error between different simulation models and guide adaptive coupling. Surveys of multi-fidelity frameworks illustrate how surrogate‐based gap estimation accelerates uncertainty quantification and optimization tasks by combining high‐ and low‐fidelity data on the fly \cite{Peherstorfer2017}. In turbulence modeling, Duraisamy et al. \cite{Duraisamy2019} review how \ac{ML}‐augmented subgrid closures can learn and quantify model-form error, demonstrating the value of data‐driven discrepancy estimation for guiding selective high‐fidelity computation. Advances in scientific \ac{ML} -- such as deep multi-fidelity \ac{GP} \cite{Raissi2016}, nonlinear information fusion in \ac{GP}s \cite{Perdikaris2017}, and time‐series \ac{ML}-error models \cite{ParishCarlberg2020} -- demonstrate the growing capability in learning and quantifying fidelity discrepancies.

Beyond methodological surveys, recent studies integrate \ac{ML} directly into multiscale simulation infrastructures. Sanderse et al. \cite{Sanderse2024} review the broad challenge of applying \ac{ML} to multiscale closure problems, emphasizing the need for physics-aware surrogate forms and discretization invariance. Nguyen et al. \cite{Nguyen2023} discuss opportunities and challenges in \ac{ML}‐augmented multiscale modeling, from data scarcity to interpretability in engineering systems. Coupling \ac{ML} surrogates with high-performance frameworks has enabled dynamic-importance sampling to select scales adaptively in real time \cite{OSTI1833796}. Additionally, Benson et al. \cite{PNAS2021} report an \ac{ML}-driven scale-bridging infrastructure for active matter simulations, which automatically learns mapping operators between fine- and coarse-scale representations.

\begin{figure}[tb!]
    \centering
    \includegraphics[width=0.6\linewidth]{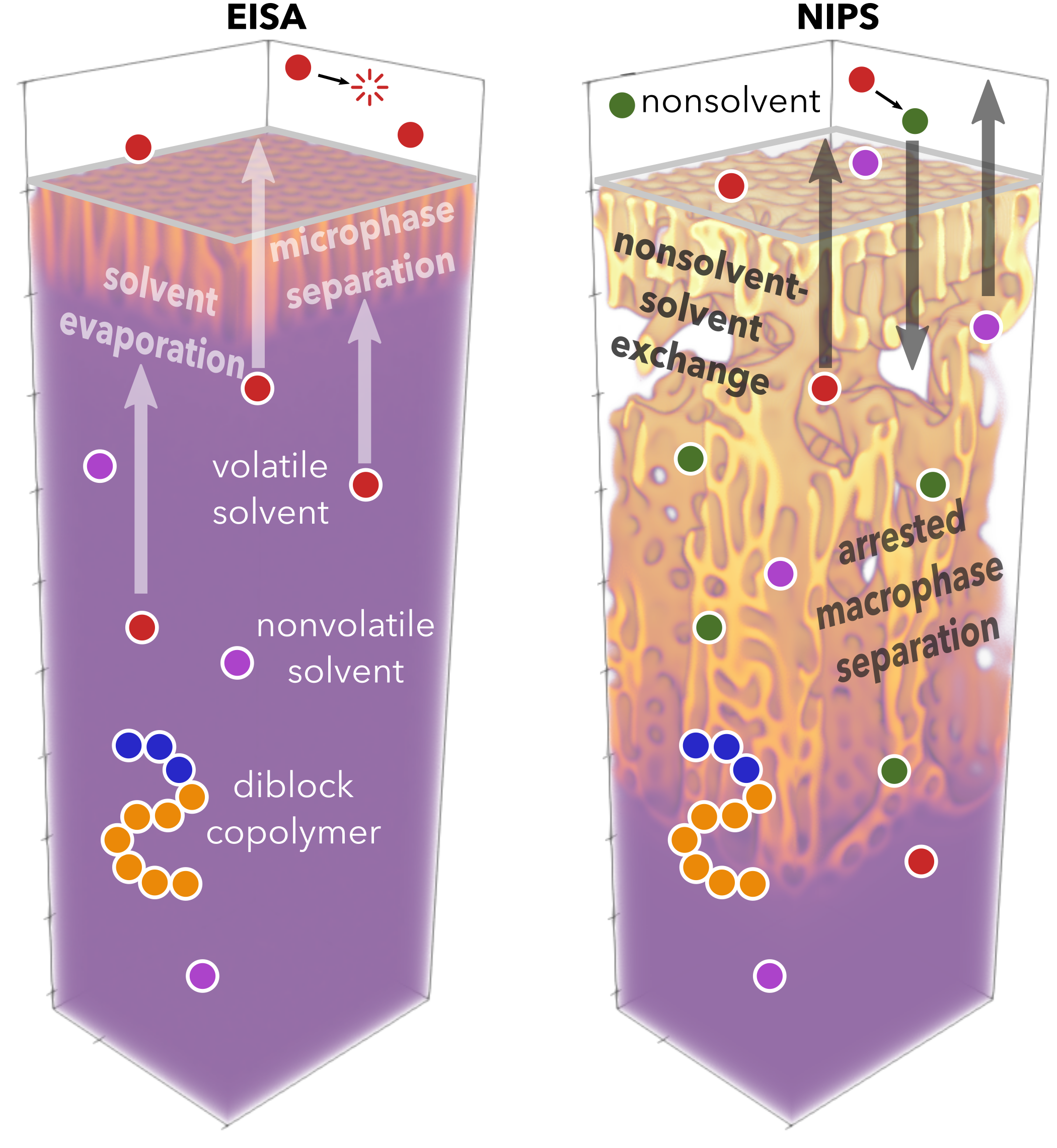}
    \caption{Schematic representation of the \acf{EISA} and \acf{NIPS} process. During \ac{EISA} the volatile solvent evaporates from the casting solution -- diblock copolymers dissolved in a two-solvents mixture -- by diffusion. The evaporation is captured in simulation by conversion of solvent into the gas phase at the top. During \ac{NIPS} nonsolvent-solvent exchange is initiated by exchanging the gas phase for nonsolvent. Inside the solution film, a combination of macrophase separation and simultaneous polymer vitrification results in the final nonequilibrium structure.}
    \label{fig:snips-scheme}
\end{figure}
In this work, we focus on the question of how to use machine learning to efficiently couple models with a different degree of details for simulating the fabrication of integral‐asymmetric, isoporous diblock‐copolymer membranes \cite{peinemann_asymmetric_2007,rangou_self-organized_2014,abetz_isoporous_2015,radjabian_tailored_2015,guo_porous_2021}. These are cast from a diblock copolymer solution and consist of an isoporous top layer, resulting in ideal selectivity, and underneath a coarse, sponge-like substructure, giving the membrane mechanical stability while retaining permeability. The fabrication process -- \ac{SNIPS} -- involves two steps: After casting a block copolymer solution onto the substrate with a doctor blade, \acf{EISA} occurs -- a volatile solvent evaporates from the film leading to the self-assembly of the diblock copolymer into a cylindrical morphology that is oriented perpendicularly to the film surface. In a second step, the liquid film is immersed in a coagulation bath -- commonly a nonsolvent such as water -- thereby inducing \acf{NIPS}. In this step, the nonsolvent penetrates the film through the cylindrical phase, leading to the formation of open pores, while at the same time, the polymer vitrifies, freezing the nonequilibrium structure, as the solvent penetrates into the film. Underneath the selective, isoporous top layer, the nonsolvent macrophase separates from the polymer, which results in the coarse sub-structure. The \ac{SNIPS} process is depicted in \autoref{fig:snips-scheme}, showing how the problem is tackeled by simulations. Achieving the ideal membrane morphology via this process is challenging because it depends on a diverse set of material and process parameters. Simulations can give valuable insights into the physics at play, as well as guide experiments by revealing parameter dependencies. With modern supercomputing, continuum and particle simulations are able to capture the relevant length and time scales, and have provided valuable insights into the \ac{EISA} process \cite{paradiso_block_2014,berezkin_vertical_2016,hao_self-assembly_2017,dreyer_simulation_2022}, the \ac{NIPS} process \cite{grzetic_modeling_2023,cooper_investigating_2024}, as well as the full \ac{SNIPS} process\cite{Blagojevic2023Dec,Blagojevic2024Oct} based on particle simulations. Nonetheless, the latter simulations are computationally demanding and large-scale parameter studies are challenging with such techniques. Continuum modeling offers a computationally more efficient approach to modeling polymer self-assembly and the \ac{SNIPS} process. Using the GPU-accelerated \ac{UDM}-software\cite{hafner_reaction-driven_2023,hafner_reaction-driven_2024}, we demonstrate that continuum modeling allows modeling the full \ac{SNIPS} process and results in qualitatively similar results as the coarse-grained particle simulations; nonetheless, quantitative differences remain.  

Planning in the future to combine the two models in a multi-fidelity modeling approach, in this work, we present an \ac{MLP}‐based decision model that predicts, at runtime, the local divergence between particle and continuum simulations. By triggering high‐fidelity simulation only where the predicted error exceeds a tolerance, a concurrent hybrid solver can be expected to achieve significant speedups while controlling accuracy -- offering a blueprint for adaptive multiscale coupling in other domains. 

\section{Simulation of membrane fabrication}

To model the two-step fabrication process, we consider a system of a cylinder-forming $AB$ diblock copolymer (typically poly(4-vinylpyridine) \linebreak (P4VP)-polystyrene (PS)) dissolved in a volatile solvent (THF) and a nonvolatile one (Tetrahydrofuran (DMF)). For the \ac{EISA}-process the gas phase is modeled as a phase-separated liquid phase, while this species is exchanged for non-solvent, water, at the start of the \ac{NIPS}-process. In the following, the different species are denoted $A$ (P4VP), $B$ (PS), $S$ (THF), $C$ (DMF), $G$ (gas) and $N$ (nonsolvent).
The two simulation schemes are shortly introduced in the following, while a more detailed description of the two is given in \ref{sec:simulation-techniques-si}.

\subsection{Particle-based simulations}

The particle simulations employ a soft, coarse-grained model \cite{daoulas_single_2006,muller_studying_2011} that represents several monomer repeat units by a single particle. The Hamiltonian is split into strong bonded (b) and weak non-bonded (nb) interactions,
$
    \mathcal{H} = \mathcal{H}_\mathrm{b} + \mathcal{H}_{\mathrm{nb}}
$.
While the bonded interactions -- harmonic springs -- are treated exactly, the non-bonded interactions are accurately replaced by interactions with fluctuating external fields. For this, the local normalized concentration fields, $\phi_\alpha(\mathbf{r})$, for $\alpha=A,B,S,C,G,N$ and space $\mathbf{r}$, and the external fields are evaluated as a function of these. The \ac{SCMF}-algorithm is employed, which transiently fixes the external fields on time scales at which they evolve only marginally and therefore decouples the molecules for a short time period. This allows for massive parallelization on modern supercomputer architectures, which is efficiently implemented in the software \ac{SOMA} \cite{schneider_multi-architecture_2019}. In the software, particle positions are propagated in time by smart-Monte-Carlo moves, resulting in Rouse-like dynamics of the polymer chains \cite{Muller2008Oct}. Hence, it is capable of modeling the non-equilibrium time evolution of the membrane fabrication process.

In the particle model, $N_0$ refers to the chain-contour discretization of the polymer and lengths are measured in units of $R_e$, the mean end-to-end distance of the ideal polymer chain. Time is measured in units of time, $\tau_R$, the time it takes the polymer to diffuse its own mean end-to-end distance. The density of the system is set by the invariant degree of polymerization, $\bar{\mathcal{N}}$, which controls the strength of thermal fluctuations. Interactions are modeled as local binary repulsions between distinct species, in the form of the Flory-Huggins parameters $\chi_{\alpha \beta}$ for $\alpha,\beta=A,B,S,C,G,N$. To model the vitrification of the polymer in the absence of plasticizing solvent, the inverse segment friction is modified depending on the local composition, $m_\alpha(\{\phi_\beta\})$. It follows a sigmoidal curve that has unity at vanishing polymer concentration and vanishes for high ones, with the turning point given by a threshold polymer concentration, $\phi_P^*$.

\subsection{Continuum simulations}

Within the continuum model, the normalized concentration fields, $\phi_\alpha(\mathbf{r})$, fully determine the system's state, and are directly propagated in type via Ginzburg-Landau-type dynamics, also known as model-B, \cite{hohenberg_theory_1977}
\begin{align}
    \frac{\partial \phi_\alpha(\mathbf{r},t)}{\partial t} = \nabla\cdot \left[R_e^2\lambda_\alpha(\{\phi_\beta(\mathbf{r}\}) \phi_\alpha(\mathbf{r}) \frac{\nabla\mu_\alpha(\mathbf{r})R_e^3}{\sqrt{\Nbar}\kT}\right] + s_\alpha(\mathbf{r}),
    \label{eq:model-B}
\end{align}
where $k_B$ is the Boltzmann constant and $T$ the temperature. The chemical potentials, $\mu_\alpha(\mathbf{r})$ are derived from the free-energy functional, $\mathcal{F}[\{\phi_\alpha\}]$, that was proposed by Uneyama and Doi for arbitrary blends of copolymers and homopolymers or solvents \cite{uneyama_density_2005,uneyama_calculation_2005}. The term $s_\alpha(\mathbf{r})$ denotes local sinks and sources that are used to model the evaporation process during \ac{EISA}\cite{dreyer_simulation_2022}, as well as the solvent-nonsolvent exchange during \ac{NIPS}.

The flexibility of the theory allows treating a wide range of systems and phenomena. In addition, the equilibrium phase behavior, as well as the nonequilibrium dynamics, shows great similarity to more detailed theories like \ac{SCFT} for polymers \cite{matsen_stable_1994,uneyama_density_2005} or the above-described particle model for a number of systems \cite{dreyer_simulation_2022,hafner_reaction-driven_2023,hafner_reaction-driven_2024}.
In \autoref{eq:model-B}, $\lambda_\alpha$ determines the mobility of each species and its variation with polymer concentration models the polymer vitrification. For this, one can simply identify the inverse segmental friction of the particle model to be $m_\alpha(\{\phi_\beta\}) = \lambda_\alpha(\{\phi_\beta\}) / \lambda_0$, where $\lambda_0^{-1}$ sets the time scale in the continuum model. 

The model implementation utilizes the \ac{UDM} software framework \cite{hafner_reaction-driven_2023,hafner_reaction-driven_2024}, which offers a CUDA/C-based GPU implementation of the aforementioned model. To model the \ac{SNIPS} process, including the required polymer vitrification, the software has been extended to feature density-dependent mobility. 

\subsection{Choice of model parameters}

Apart from the time scale, both the continuum and the particle model operate on the same set of parameters, which enables an excellent comparison between the two models. As the base parameter set of this work, we employ the same parameters as established in the work of Blagojevic \textit{et al.} \cite{Blagojevic2024Oct}, for which the \ac{SNIPS} process results in an ideal membrane morphology within the particle simulations. Deviations from the base parameter set will be specifically denoted. Most importantly, the interactions are varied for the generation of training data. The process and material as well as discretization parameters are chosen as follows.

In the particle model, the chain-contour discretization of the diblock copolymer is chosen $N_0=64$, while solvent molecules are modeled as \linebreak oligomers of $N_\alpha=8$ for $\alpha=S,C,G,N$. In the continuum model, only the ratio of the two plays a role which is kept the same. The block ratio of the diblock copolymer is chosen $f_A=0.3125$, and in the particle model, we choose $\sqrt{\bar{\mathcal{N}}}=380$. The symmetric matrix of Flory-Huggins parameters for the reference system reads

\begin{align}
\chi N_0 = 
\begin{pmatrix}
 & A & B & S & C & G & N \\
A & 0 & 30 & 15 & 0 & 200 & 10 \\
B & 30 & 0 & 0 & 3 & 140 & 150 \\
S & 15 & 0 & 0 & -35 & 10 & -30 \\
C & 0 & 3 & -35 & 0 & 120 & -30 \\
G & 200 & 140 & 10 & 120 & 0 & - \\
N & 10 & 150 & -30 & -30 & - & 0
\end{pmatrix}.
\end{align}
These are inspired by experimental findings and chosen such that the volatile solvent, $S$, is selective for the matrix-forming block $B$, whereas the nonvolatile solvent, $C$, prefers the cylinder-forming block, $A$. The nonsolvent, $N$, is highly incompatible with the polymer, particularly with the matrix-forming block, $B$, leading to macrophase separation upon penetration into the solution film. For the generation of training data, we wish to cover a broad range of scenarios within the \ac{SNIPS} process. Therefore, we randomly vary this interaction matrix, as will be described in detail below.

Finally, to make the simulations fully comparable, the time scales, $\lambda_0^{-1}$ in the continuum model and $\tau_R=31\,355$MCS (Monte-Carlo steps) in the particle model need to be matched. For this study, the ratio of $\tau_R\lambda_0 = 1.3125$ was determined by minimizing the deviation of concentration fields during the \ac{EISA} process.

In the continuum model, a time step of $\Delta t = 2\cdot 10^{-5}\lambda_0^{-1}$ is chosen to propagate the system in time. 

The parametrization of the species mobility is chosen, such that the threshold polymer concentration is $\phi_P^*\approx 0.72$. The initial casting solution is disordered with equal concentrations of polymer and the two solvents $\phi_P(t=0)=\phi_S(t=0)=\phi_C(t=0)=\frac{1}{3}$. For the introduction of the modeling, we choose domains of size $V=14 \times 16 \times 51.2 R_e^3$, which is discretized into a collocation grid with spacing $\Delta x=\frac{R_e}{10}$. For the generation of training data, a smaller domain, with size $V=5.6 \times 6.4 \times 51.2 R_e^3$, was chosen.  In all cases, the domain has periodic boundary conditions in $x$- and $y$-direction, and a no-flux boundary condition in $z$-direction. This is achieved via reflective boundary conditions in the continuum model, and in the particle model via a wall, which particles cannot penetrate. The process time for \ac{EISA} and \ac{NIPS} are chosen equal for both, $T_\mathrm{EISA/NIPS} = 16\tau_R = 21 \lambda_0^{-1}$ in the particle and continuum simulations respectively.

\subsection{Example simulations}
\label{sec:membrane_fabrication}

\begin{figure}[t!]
    \centering
    \includegraphics[width=\textwidth]{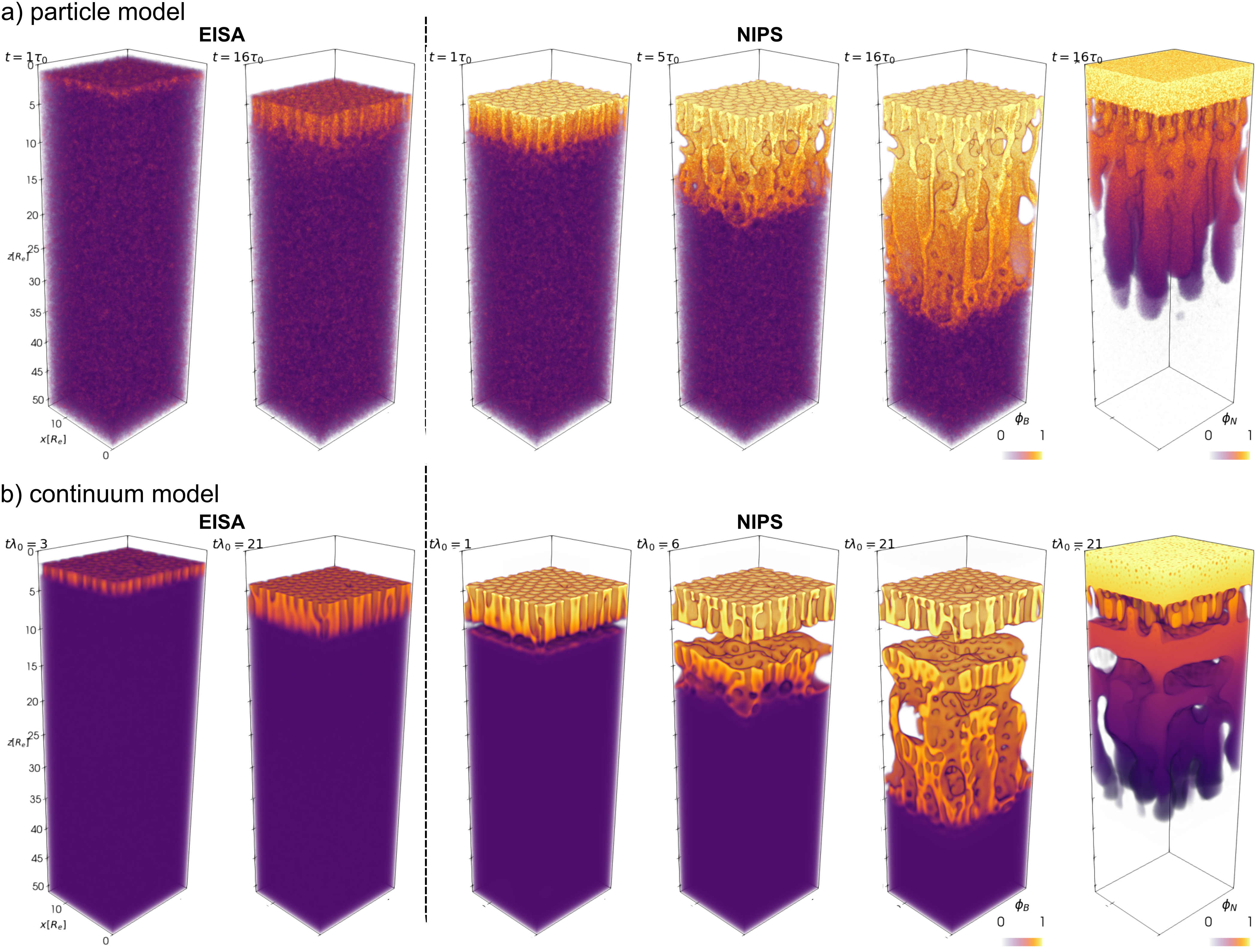}
    \caption{Time evolution of the B-block concentration fields, $\phi_B(\mathbf{r})$, during the \ac{EISA} and \ac{NIPS} process in a) the particle simulations and b) the continuum simulation. Indicated times are measured from the start of each process. The far right panel additionally shows the nonsolvent concentration field, $\phi_N(\mathbf{r})$, after the full \ac{SNIPS} process.}
    \label{fig:time-evolution-example}
\end{figure}
With this setup, we are able to model the \ac{SNIPS} process with both simulation schemes. We present the complete \ac{SNIPS} process for both simulation approaches using the reference system parameters. \autoref{fig:time-evolution-example} shows 3D concentration fields of the matrix-forming block B during membrane formation at different times for the two simulation schemes, as well as the final concentration field of the nonsolvent, $\phi_N(\mathbf{r})$. 

During \ac{EISA} ($t < 16\tau_R$ in the particle model, $t<21\lambda_0^{-1}$ in the continuum model), the volatile solvent, $S$, evaporates, causing the film surface to retract and the polymer to enrich at the interface. This increased polymer concentration at the surface -- skin formation -- promotes its self-assembly into hexagonally ordered cylinders perpendicular to the film surface, forming a well-ordered layer. After the solvent-evaporation step, this layer exhibits a thickness of approximately $4R_e$. These dynamics are equivalent in both simulation schemes. 

Upon the next process step, at time $t=t_{\text{EISA}} = 16\tau_R = 21\lambda_0^{-1}$, \ac{NIPS} initiates through an instantaneous conversion of gas to nonsolvent. The nonsolvent, $N$, being incompatible with polymer but miscible with solvents S and C, penetrates the polymer skin primarily through the A-rich cylinder cores due to preferential compatibility. As remaining solvent is displaced, the polymer concentration exceeds the vitrification threshold $\phi_P^*$, arresting the self-assembled structure in a glassy state.

\begin{figure}[tbp!]
    \centering
    \includegraphics[width=0.9\textwidth]{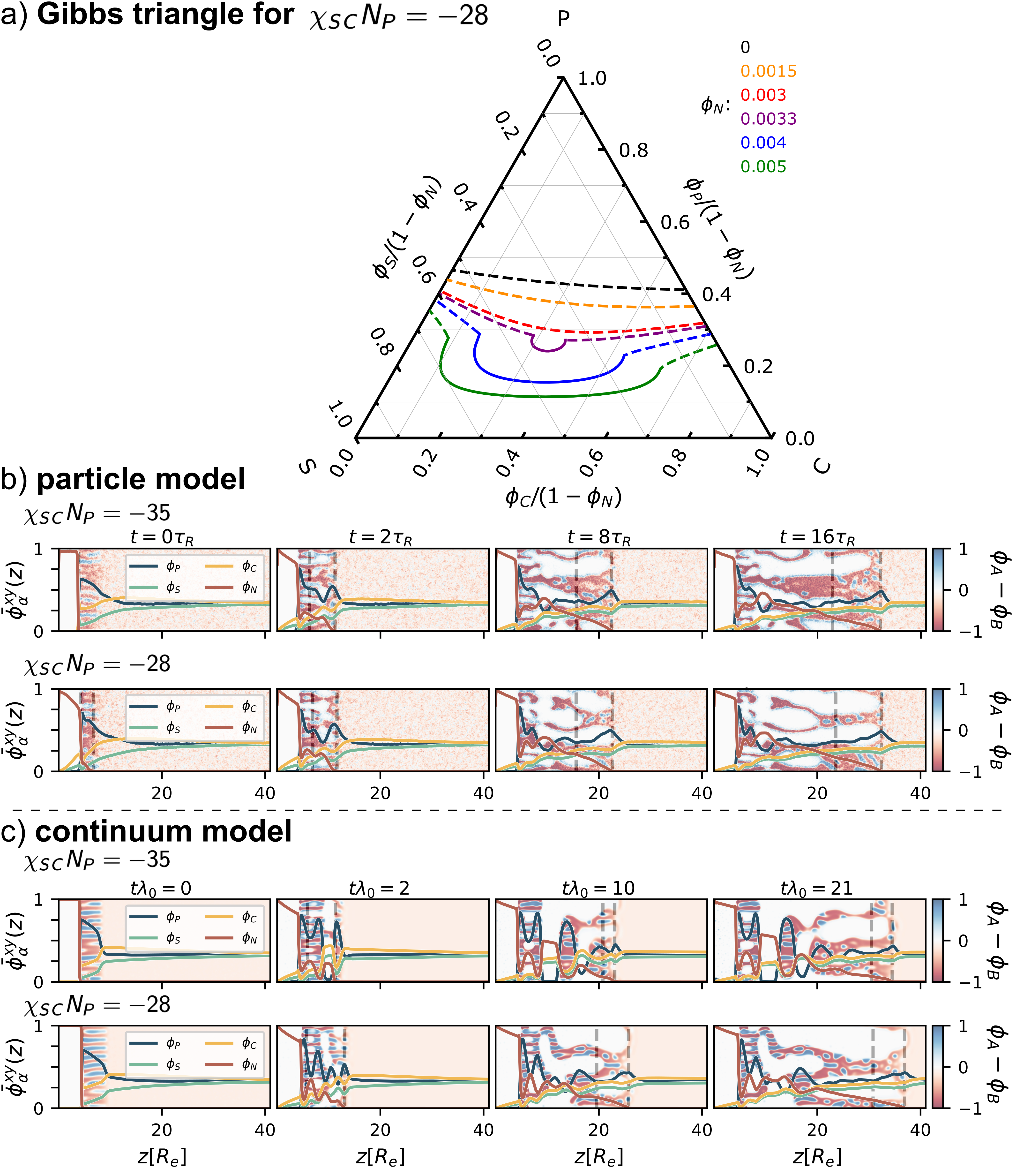}
    \caption{Panel a) shows spinodal curves of membrane casting solutions at different nonsolvent concentrations on the Gibbs triangle. Dashed and solid lines indicate instabilities of the spatially homogeneous state against composition fluctuations with finite and infinite length scales, respectively. Panels b) and c) show the  1D laterally-averaged concentration profiles, $\bar{\phi}_\alpha^{xy}(z)$ (lines) and 2D cross-sections of the block concentration difference, $\phi_A(\mathbf{r})-\phi_B(\mathbf{r})$ (background) for a) the particle simulations and b) the continuum simulations of the base parameter pair (top) and the weakened solvent attraction, $\chi_{SC}N_P=-28$ (bottom). Time increases from left to right as indicated. For the 2D cross sections, the $y$-axis is scaled the same as the $z$-axis. The approximated front positions are indicated as dashed gray lines -- vitrification front left of the structure-formation front. 
    }
    \label{fig:lateral-average-comparison}
\end{figure}

The thermodynamics underpinning structure formation during \ac{EISA} and \ac{NIPS} can be understood from the spinodal curves of membrane casting solutions at different nonsolvent concentrations \cite{Blagojevic2024Oct}, shown on the Gibbs triangle in \autoref{fig:lateral-average-comparison}(a). The spinodal, evaluated using the random-phase approximation (RPA)\cite{de1979scaling, leibler1980theory}, marks the loss of linear stability of the spatially homogeneous state against composition fluctuations with either a finite (dashed lines) or infinite (solid lines) characteristic length scale. As the RPA is a standard approach for determining the spinodal in multicomponent polymer systems, we refer the reader to the literature for technical details \cite{leibler1980theory, hong1983theory, whitmore1985theory, xie2024phase}. The spinodal curve with $\phi_{N}=0$ delineates the instability boundary of the casting solution used in \ac{EISA}. At $t=0$, the initial concentrations, $\phi_P(t=0)=\phi_S(t=0)=\phi_C(t=0)=\tfrac{1}{3}$, lie within the (meta)stable region of the Gibbs triangle. As solvent S evaporates, $\phi_S$ decreases near the solution surface while $\phi_P$ and $\phi_C$ increase, shifting the state point of this layer toward the spinodal (black dashed curve in \autoref{fig:lateral-average-comparison}(a)). Upon crossing the spinodal, spontaneous phase separation occurs at a finite length scale, giving rise to the vertically arranged cylinders observed in \autoref{fig:time-evolution-example}.

During \ac{NIPS}, nonsolvent molecules penetrate the film as solvent and nonsolvent exchange. As shown in \autoref{fig:lateral-average-comparison}(a), the (meta)stable region of the solution shrinks with increasing $\phi_N$, driven by the strong polymer–nonsolvent incompatibility. Notably, once $\phi_N$ exceeds a threshold of about $\phi_N^\ast=0.0033$, a segment of the spinodal curve corresponding to instability against fluctuations of infinite length scale emerges (solid line). This solid instability boundary separates the dashed boundary into two branches and intersects them at two points, i.e., the Lifshitz points \cite{xie2024phase, xie2025topological}. In contrast to the dashed portions of the spinodal curve, the solid segment expands much more rapidly with increasing $\phi_N$ and dominates the entire spinodal for $\phi_N \gtrsim 0.004$. When the state point of the casting solution crosses this solid boundary, phase separation between polymer and nonsolvent proceeds on a macroscopic length scale, giving rise to the macroporous structure observed in \autoref{fig:time-evolution-example}. 

The two simulation approaches exhibit qualitatively different behavior during \ac{NIPS}. In the particle simulations, nonsolvent penetration induces controlled macrophase separation, leading to the formation of a sponge-like substructure with moderate-sized macrovoids connected to the cylindrical top layer. In contrast, the continuum model predicts more pronounced macrophase separation upon nonsolvent penetration, resulting in the formation of large macrovoids spanning the entire lateral simulation domain. This leads to complete detachment of the sponge-like substructure from the cylindrical open-pore top layer, creating a distinct interface between these regions. The continuum model's prediction of complete layer detachment represents a more extreme manifestation of the phase separation instabilities, while the particle-based approach captures the intermediate connectivity observed experimentally in many \ac{SNIPS} membranes.

This difference of dynamics during the \ac{NIPS} process can also be observed in \autoref{fig:lateral-average-comparison} b) and c), which show the laterally averaged concentration profiles of each species, as well as a 2D slice of the concentration difference between the two polymer blocks, $\phi_A-\phi_B$, at different times during the \ac{SNIPS} process in both models. In each panel, the top row shows the base parameter combination. The detachment of the cylindrical phase from the substructure in the continuum simulation expresses itself in a vanishing polymer concentration profile at $z\approx 10 R_e$. In contrast, the polymer concentration stays finite in this region in the particle simulation. Nonetheless, a dip of this concentration is also visible. 

A minor change to the base parameter combination changes the membrane morphology qualitatively in the continuum simulation, resulting in the connectivity of the two membrane regions. This occurs when the $S$- and $C$-solvent attraction that leads to a depletion of solvents in the polymer skin, is decreased, $\chi_{SC}N_0=-28$. This scenario is plotted in the bottom row of each panel in \autoref{fig:lateral-average-comparison} b) and c), which shows that the decreased attraction leads to weaker concentration gradients inside of the cylindrical domains after \ac{EISA}, so that the nonsolvent penetration can be expected to be slower after initiation of \ac{NIPS}. Additionally, the less pronounced increase of solvent concentration underneath the polymer skin provides a reduced pre-pattern for the macroscopic nonsolvent phase to form in this region. In contrast, in the particle simulation the change in material parameters makes no qualitative difference. 

\begin{figure}
    \centering
    \includegraphics[width=0.5\linewidth]{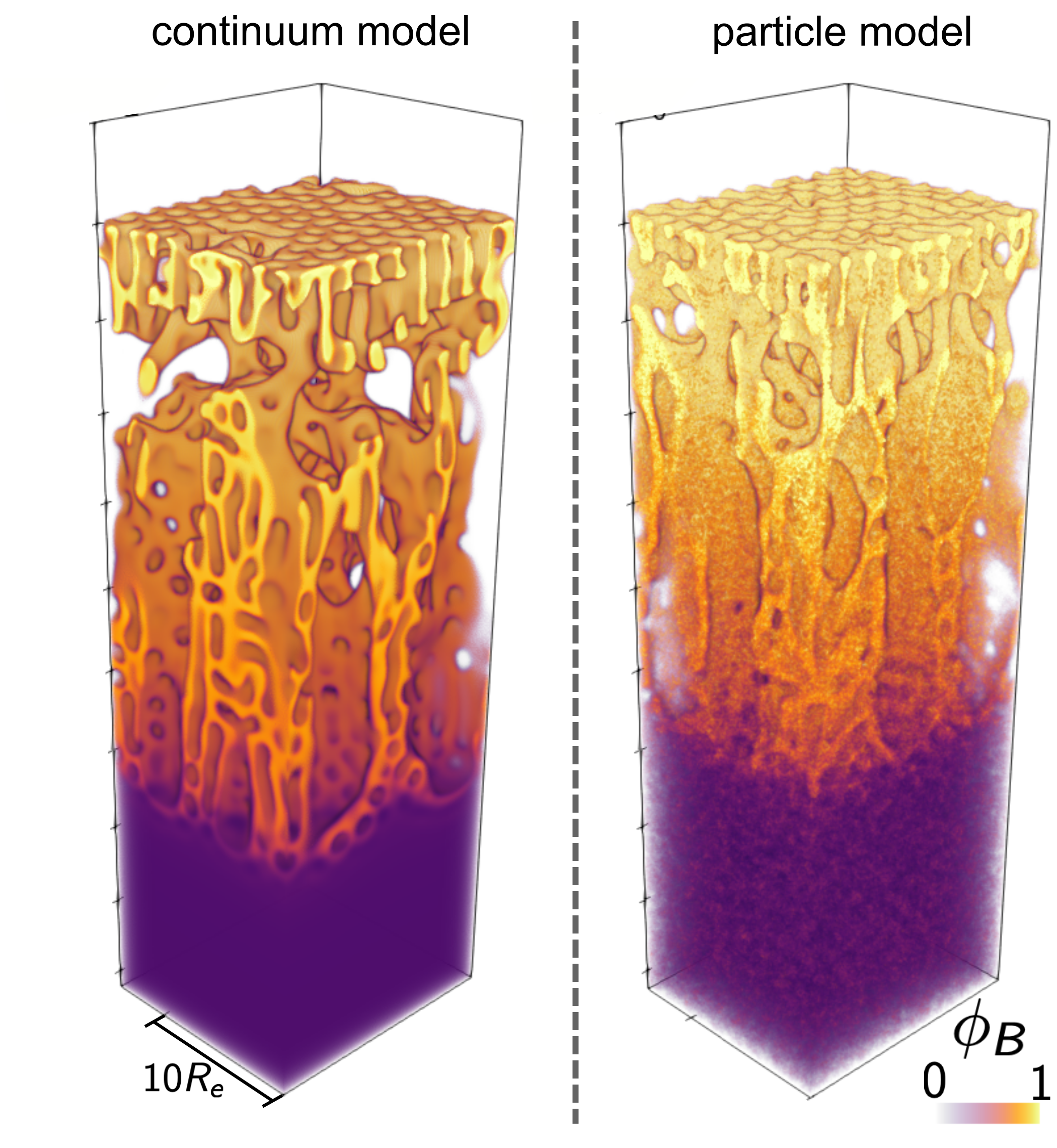}
    \caption{Final $B$-block concentration field, $\phi_B(\mathbf{r})$ after the \ac{NIPS} process in the continuum simulation (left) and in the particle simulation (right) for the reduced solvent attraction, $\chi_{SC}N_P=-28$, compared to the base parameter combination.}
    \label{fig:improved-parameters-morphology}
\end{figure}
The final 3D $B$-block concentration field after \ac{NIPS} for the solvent attraction is shown in \autoref{fig:improved-parameters-morphology}, showing the polymer structure has a hexagonal top-layer that is connected to the sponge-like substructure underneath in both models. Hence, the continuum simulations are qualitatively giving similar results, with variations in the exact parameters values, but need to be adjusted to give a full quantitative overlap with the more-accurate particle simulations.

In addition to the morphological analysis, the two simulation approaches reveal a three-layer structure during the later stages of \ac{NIPS} ($t > 36\tau_R$), that effectively describes a 1D problem:

\textbf{Layer 1: Arrested phase separation.} This vitrified region extends from the hexagonal polymer skin at the film surface into the upper macroporous substructure. Here, low solvent concentrations in polymer-rich domains cause vitrification of the matrix-forming block B upon crossing the glass-transition threshold. Dynamics consist primarily of nonsolvent-solvent exchange through the static, complex porous geometry.

\textbf{Layer 2: Active structure formation.} This region extends from the vitrification front to the structure-formation front, where nonsolvent concentration sufficiently high to induce macrophase separation and/or solvent depletion triggers microphase separation. The macro- and microphase separation fronts coincide for the chosen \ac{SNIPS} parameters, resulting in polymer concentration increase and A-core micelle formation.

\textbf{Layer 3: Homogeneous solution.} Deep within the film, the solution remains laterally homogeneous without phase separation. Only shallow concentration gradients exist.
%, with dynamics governed by one-dimensional solvent-nonsolvent exchange through the unstructured medium.
%\MM{In general, this is true but in this specific case, I would rather say that the solvent(s) gradient are compensated by a 1D polymer-density variation (increase at the structure-formation front), but there is no nonsolvent -- a tiny amount of nonsolvent would already result in phase separation. Maybe we only say that -- Only shallow concentration gradients exist.}

The transitions between these three layers, the vitrification front and the structure-formation front, are approximated for all simulations and indicated in \autoref{fig:lateral-average-comparison}. To approximate the vitrification front, we take the position where laterally-averaged absolute difference of the current concentration with the future concentration, $\Delta \tau=4.2\lambda_0^{-1}=3.2\tau_R$, falls below a $30\%$ threshold of the maximal change. Hence this quantifies the region in which the future concentration change is sufficiently small. This results in the left gray line of the panels in \autoref{fig:lateral-average-comparison} b) and c). To approximate the structure-formation front, we use the Gibbs triangle of \autoref{fig:lateral-average-comparison} a), and identify the lowest nonsolvent concentration for which macrophase separation occurs to be $\phi_N=0.0033$. Hence, the structure-formation front occurs at the largest $z$-value, where the laterally-averaged nonsolvent concentration crosses this critical concentrations, and is observable as the right gray line in the panels of \autoref{fig:lateral-average-comparison} b) and c). With these fronts indicated, the above-described three layers become clearly visible. 

Improvement of the continuum simulations most likely target layer 2, where the structure formation occurs. These are therefore spatially confined, hinting towards our planned strategy of coupling the full continuum simulations to locally-confined, and therefore computationally more efficient, particle simulations.

\subsection{Strategy for efficient SNIPS simulations}

Given the qualitative similarity of the continuum model results compared with the more accurate particle simulations, the continuum simulations appears ideal to perform high-throughput parameter studies of the \ac{SNIPS} process. While the qualitative similarity is given, there are clear quantitative differences that need to be overcome.

The above-described three-layer model
provides a strategy for using both models to perform a computationally 'efficient, yet accurate simulation. The main differences between the two simulation schemes can be expected to occur in the second layer -- the structure formation zone. Performing a small-scale particle simulation in this narrow zone, across the full lateral dimensions, while modeling the full domain with the continuum model, would provide the required accuracy during the structure formation. Such a scheme remains computationally efficient because the particle simulation is only performed on a small sub-domain.

However, identifying the size and position of the sub-domain requires \textit{a-priori} knowledge, where the continuum model deviates most from the particle model. In the following, we demonstrate that this can be achieved through a \ac{MLP}-based prediction of the laterally-averaged error. The choice of sub-domain can then be performed by a simple post-processing of the live-performed error prediction. This \ac{ML}-based decision model should work reliably for a wide range of parameters, preferably even outside of the range of the training data. Here, we can make use of the fact that the \ac{SNIPS} process is expected to follow the above three-layer model. Therefore, the \ac{ML}-model should be able to predict the ideal sub-domain even if it is only trained on a finite range of interaction parameters.

% \begin{itemize}
%     \item Example simulation results of SNIPS with both models
%     \item Describe physics at play during SNIPS (ternary phase diagram, 1D three-layer mode (1. homogeneous 2. structure-formation 3. frozen) $\rightarrow$ serves to justify choice of 1D error prediction, compare/describe 1d-diffusion models
%     \item Clarify qualitative similarity of results but show quantitative differences
%     \item Identify (dis-)advantages of each model --> need to identify where continuum model deviates from particle simulations
%     \item what is expected from \ac{ML}-based error prediction/ sub-domain choice (Potentially separate sub-section)
% \end{itemize}

\section{Decision-model architecture}
% Explain architecture of decision model: preproceesing, MLP application and postprocessing and how the data streams flow, show flow chart 
\begin{figure}[t!]
    \centering
    \includegraphics[width=0.72\textwidth]{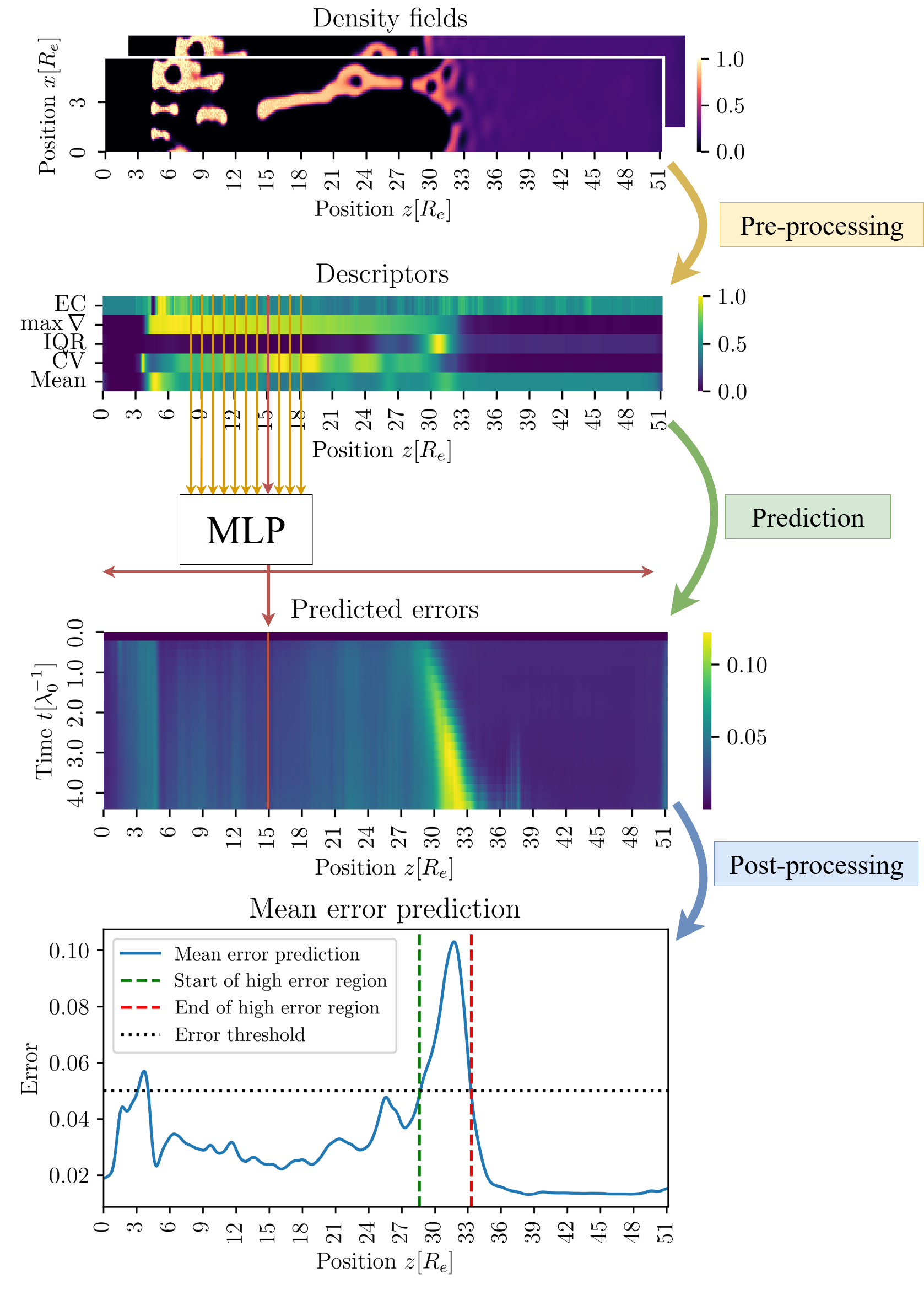}
    \caption{Overview diagram for the decision model. The different steps of data processing with intermediate results are shown. For details on each step, refer to the following sections \ref{preprocessing}, \ref{prediction} and \ref{postprocessing}. In the pre-processing step descriptors are extracted from the density field (5 for each $z$-layer, abbreviations: CV -- coefficient of variation; IQR -- interquartile range of finite differences; max$\nabla$ -- maximum gradient magnitude; Mean -- layer mean; EC -- Euler characteristic)%\MM{for two consecutive time steps}
    , with the IQR being computed from the finite difference of two consecutive (in time) density fields. The prediction is done by a \ac{MLP}, which takes descriptors of multiple $z$-layers as input and outputs predicted errors for a single $z$-layer at multiple future time steps. The post-processing aggregates the errors for the requested time range and finds the largest error region above a threshold. }
    \label{fig:decision_model_architecture}
\end{figure}

An overview of the architecture of the decision model for the NIPS-process is shown in \autoref{fig:decision_model_architecture}. The model has to start with a raw density field from the continuum simulation and arrive at a partitioning for the domain in the end. The process consists of the following three steps:%\GH{Should we make these subsubsection or should we go back to separate methods section for simulations and ML? -- did the latter -- OK? Yes}
\begin{itemize}
    %\item \textbf{Production of training data(\autoref{sec:training-data}}: Simulations are performed with both models for a broad range of process parameters and from the same initial conditions, starting from all stages of the fabrication process.
    \item \textbf{Pre-processing (section \ref{preprocessing}):} The raw data, the B-block concentration field $\phi_B(\mathbf{r})$, is smoothed  with a Gaussian filter and information-dense descriptors are extracted. The descriptors reduce the dimensionality of the data by two, such that only the vertical dimension, $z$, remains.
    \item \textbf{Prediction (section \ref{prediction}):} A trained machine learning model, an \acf{MLP}, estimates the time evolution of $z$-resolved divergence between both simulation methods.
    \item \textbf{Post-processing (section \ref{postprocessing}):} The divergence is interpreted as the local error, where a large divergence means that the continuum simulation is not an accurate approximation and the particle simulation needs to be applied. 
\end{itemize}

This architecture aims at providing an efficient, flexible, and reliable decision model to partition the domain based on an error prediction. The flexibility denotes the variability of domain size in all three dimensions, independence from simulation time, as well as the ability to predict the divergence for a broad range of parameter combinations. Variations of the latter may change the behavior of the simulation significantly. Within all scenarios, the model needs to reliably predict at least qualitatively correct spatiotemporal regions in which the continuum simulations are performing at an accuracy below a prescribed threshold. In the next sections, details of the architecture will be explained. In this work, we focus on the \ac{NIPS} process. However, the procedure is the same for the EISA process and the resulting \ac{MLP} differs only in the selection of descriptors and the input phase.

\subsection{Pre-processing}
\label{preprocessing}
The aim of the pre-processing is to provide reliable, information-dense input to the machine learning model while reducing data complexity. In our case, the steps include Gaussian smoothing, finite-difference computation, descriptor extraction, and scaling. 

\subsubsection*{Raw data}
The concentration fields of the five different species contain mostly redundant information. To increase efficiency, only the B-block concentration field, corresponding to the matrix-forming membrane material, is used as input. Hence, the raw data input consists of the B-block concentration fields at two consecutive time steps, $\phi_B(\mathbf{r},t)$ and $\phi_B(\mathbf{r},t-0.21\lambda_0^{-1})$. The two consecutive time steps are used to calculate the finite temporal difference, providing both the current density field and its temporal change as inputs for the descriptor calculation (\autoref{fig:decision_model_architecture}).

% \subsubsection*{Domain characteristics guiding feature extraction}
% \label{prerequisites}
% The selected phase is represented by a 3D density field describing the membrane’s structure formation process. In this process, a \textit{structure-formation front} separates a homogeneous fluid region (high $z$) from the formed membrane structure (low $z$). Initially near $z \approx 0$, the front propagates towards increasing $z$ during the simulation (\autoref{fig:decision_model_architecture}).  

% These structural characteristics motivate our domain partitioning and dimensionality reduction strategy:
% \begin{itemize}
%     \item \textbf{Localized high-dynamics region:} The structure-formation front contains the highest rates of change, while the rest of the domain evolves much more slowly. This makes it possible to simulate the front region with a high-accuracy model and approximate the rest with a lower-cost model.
%     \item \textbf{Axial invariance:} The membrane grows along the $z$-axis, and properties remain approximately constant in orthogonal directions. This allows the reduction of each 3D density field to a sequence of 2D $z$-layers, each represented by a compact set of descriptors (\autoref{fig:decision_model_architecture}).
% \end{itemize}

\subsubsection*{Descriptors}
\begin{figure}[ht]
    \centering
    \begin{subfigure}{0.48\textwidth}
        \includegraphics[width=\linewidth]{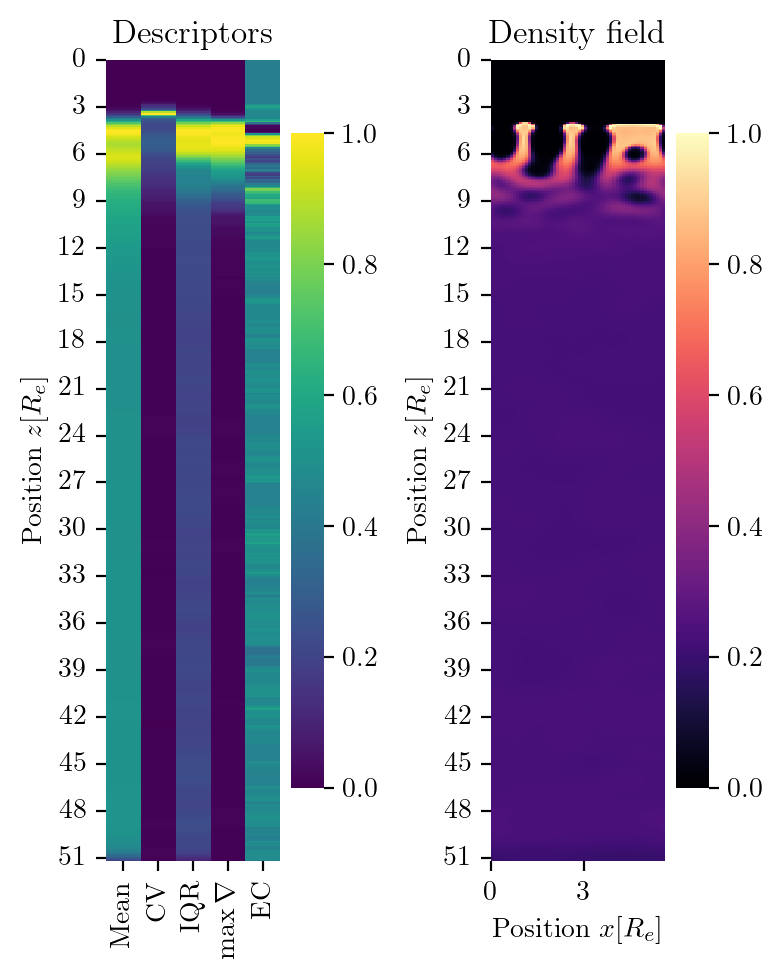}
        \caption{Selected descriptors at $T=0\lambda_0^{-1}$}
    \end{subfigure}
    \hfill
    \begin{subfigure}{0.48\textwidth}
        \includegraphics[width=\linewidth]{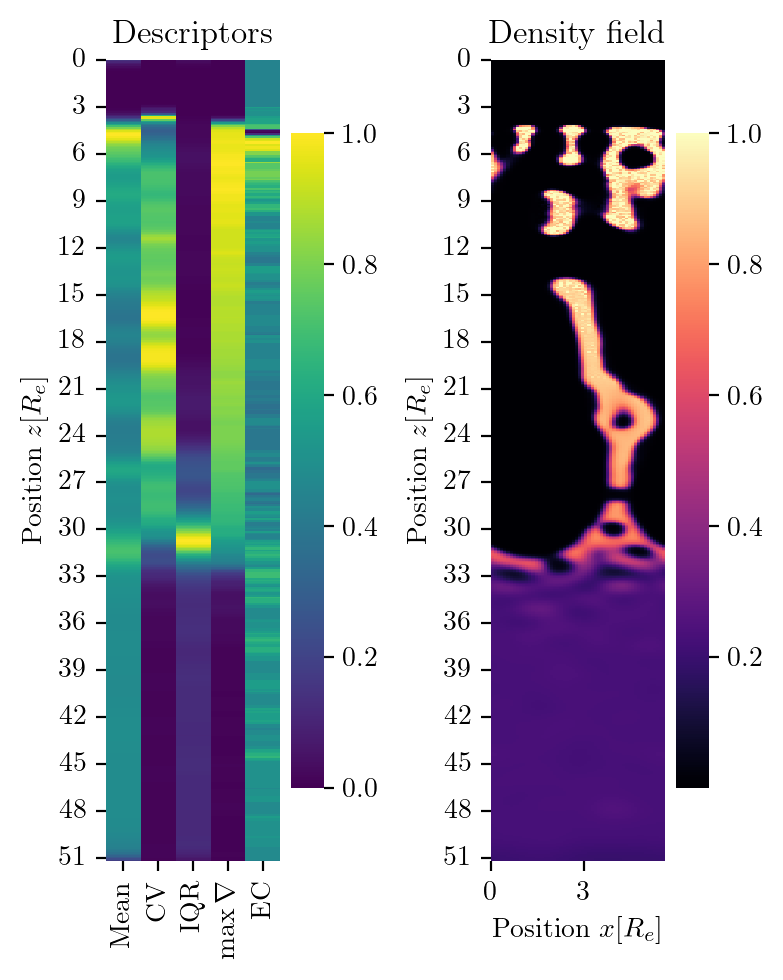}
        \caption{Selected descriptors at $T=8.4\lambda_0^{-1}$}
    \end{subfigure}
    \caption{Heatmaps showing the descriptors selected by the greedy feature selection compared to a 2D slice through the density field, $\phi_B(\mathbf{r})$, they were extracted from. On the $x$-axis the selected descriptors (abbreviations: CV -- coefficient of variation; IQR -- interquartile range of finite differences; max$\nabla$ -- maximum gradient magnitude; Mean -- layer mean; EC -- Euler characteristic) and the $x$ position are shown, while the $y$-axis depicts the $z$-position and is shared between all diagrams. The descriptors are calculated for each $z$-layer of the density field.}
    \label{fig:descriptors}
\end{figure}
A total of 63 descriptors were computed for each $z$-layer (spacing $0.1 R_e$) of the density field, $\phi_B(\mathbf{r})$, and its temporal finite difference in every sample, including statistical measures (e.g., mean, variance, percentiles) and image-based features that treat the density field as a grayscale image \cite{AwadHassaballah2016}. From these, a greedy feature selection algorithm identified the five best performing features. It started with a single feature and iteratively chose the respective best performing next feature. After five features, the improvements became negligible, and the contribution of the fifth feature was already unclear (see \autoref{fig:shap}). %Details can be found in \autoref{Asec:descriptors}.
%\MM{a more explicit explanation is needed}
%These capture complementary aspects of the evolving density field, including statistical variation, structural gradients, large-scale average intensity, and topological complexity.

%From this set, a greedy feature selection algorithm identified the \textbf{five most informative descriptors} for predicting the divergence field. 
The five best performing descriptors for prediction are the coefficient of variation (CV), the interquartile range (IQR) of finite differences, the maximum gradient magnitude, the mean, and the Euler characteristic. This section defines these descriptors, explains how they are computed, and describes their physical interpretation.

\begin{description}
  \item[Mean $\mu$] \hfill \\
  The mean $\mu$ of the layer describes changes in the structure, similar to the CV and marks the beginning and the end of the structure. However, compared to most other descriptors, there is no significant decrease to zero after the structure-formation front:
  \begin{equation}
    \mu = \frac{1}{N_x N_y} \sum_{i=1}^{N_x} \sum_{j=1}^{N_y} \phi_B(i\Delta x, j\Delta y, z)
  \end{equation}
  where $N_x, N_y$ are the number of grid cells in $x$- and $y$-direction.

  \item[Coefficient of variation (CV)]\hfill \\
  The CV of the 2D layer correlates strongly with the formed structures in the domain and is defined as
  \begin{equation}
    \text{CV} = \frac{\sigma_{\text{layer}}(z)}{|\mu_{\text{layer}}(z)| + \epsilon}
  \end{equation}
  where $\mu_{\text{layer}}(z)$ and $\sigma_{\text{layer}}(z)$ refer to the laterally (over $x$- and $y$-\linebreak dimension) averaged mean and standard deviation of the concentration field $\phi_B(\mathbf{r})$. $\epsilon = 10^{-10}$ is needed to avoid division by zero.

  \item[Interquartile range (IQR) of finite difference]\hfill \\
  The IQR of the finite difference layer measures the spread of the middle $50\%$ of values. Because it is calculated from the finite difference of two consecutive (in time) fields, it correlates strongly with the position of the structure-formation front. It exhibits a single peak at this front and is close to zero elsewhere. 
  \begin{equation}
    \text{IQR} = Q_{0.75} - Q_{0.25}
  \end{equation}
  where $Q_{0.75}$ and $Q_{0.25}$ are the 75th and 25th percentiles of the concen-\linebreak tration-field finite difference. The percentile $Q_p$ represents the value below which $p\%$ of the data points lie, providing a general measure of the data distribution.
  %\MM{This description is not clear to me -- a formula would help} \MB{A formula for the percentiles (there is no nice formula) or the finite difference ($\phi_B(r_{i,t})-\phi_B(r_{i,t-1})$, there is the timestep division left out because it must be the same)? }

  \item[Maximum gradient magnitude ($\max \nabla$)]\hfill \\
  The maximum gradient magnitude in the layer indicates structures. This feature provides a very smooth transition from the top of the solution film to the structure-formation front. With this behavior it provides a reliable indicator of the position within the structure.
  \begin{equation}
    \max \nabla = \max_{xy} \left( \sqrt{ \left( \frac{\partial \phi_B(\mathbf{r})}{\partial x} \right)^2 + \left( \frac{\partial \phi_B(\mathbf{r})}{\partial y} \right)^2 } \right)
  \end{equation}

  \item[Euler characteristic (EC)] \hfill \\
  The Euler characteristic describes the difference between the number of components and the number of holes in the binarized layer. The threshold for binarization is at $\min_{xy}(\phi_B) + 0.75 (\max_{xy}(\phi_B)-\min_{xy}(\phi_B))$. 
  %\MM{Is the last $\min$ also over $x$ and $y$? -- then it is the same as the first term. How is this (complicated) formula motivated?} \MB{You want to have the threshold somewhere between min and max and with this value (0.75) you can vary the exact position. In the optimization process there were also the 0.25 and 0.5 variants.} 
  This descriptor shows deviations over the built structure. However, because it is very noisy, its benefit to the prediction remains unclear.
  \begin{equation}
    \text{EC} = \frac{1}{L_xL_y}\left(C_{0.75} - H_{0.75}\right)
  \end{equation}
  where $C_{0.75}$ is the number of connected components and $H_{0.75}$ is the number of holes in the binarized layer. $L_x$ and $L_y$ denote the size of the layer and are used for scaling the descriptor.
\end{description}

\autoref{fig:descriptors} shows the spatial distribution of the five selected descriptors for an example simulation at two time steps, alongside the corresponding density field slices. Descriptors are plotted along the \(z\)-axis, with the descriptor index and \(x\)-position on the abscissa. This representation illustrates how different descriptors capture distinct aspects of the evolving density field. All descriptors are domain size independent, meaning that they do not need to be scaled with varying domain sizes. Specifically in the case of the Euler characteristic, the holes in the domain are counted, scaling linearly with the cross-sectional area, $L_xL_y$, of the domain, which justifies inversely scaling with it.

%\autoref{fig:descriptors} illustrates the spatial distribution of the selected descriptors for an example simulation at two time steps, alongside the corresponding density slices. The descriptors reveal clear structural patterns, such as the location of the structure-formation front and the transitions between distinct structural regimes. A full description and analysis of the selected descriptors, their definitions, and the feature selection procedure are provided in the Supplementary Information.

\subsection{Error Prediction}
\label{prediction}
% explain the training data and the assumptions underlying them, especially that it was trained with pure SOMA and UDM simulations, assuming that a coupled simulation would behave similar. Refer to SI for hyperopt
% show error prediction results vs correct errors (like in report)
\subsubsection*{Machine learning model}
The goal of the machine-learning model is to predict the layer errors for multiple time steps. With the layer being indexed by $l$ and the time being $t$, the error is defined as the mean error of the respective layer:
\begin{equation}
    \text{Error}(t,l) = \frac{1}{N_xN_y} \sum_{i=1}^{N_x}\sum_{j=1}^{N_y} \left|\phi_B^\mathrm{UDM}(\mathbf{r}_{ijl},t)-\phi_B^\mathrm{SOMA}(\mathbf{r}_{ijl},t)\right|
\end{equation}
This task is satisfactorily achieved by an \acf{MLP}, which was implemented using scikit-learn \cite{scikit-learn}. It provides an easy-to-use, reliable, and efficient implementation of a neural network with all necessary flexibility for our use case. The \ac{MLP}'s error prediction process, which it does on a layer-by-layer basis, is visualized in \autoref{fig:decision_model_architecture}. It predicts the errors of the next 21 time steps at once via its 21 outputs. Iterating the \ac{MLP} through the whole $z$-dimension, one obtains a heatmap as shown in \autoref{fig:decision_model_architecture}. It shows an estimate for the future error at each layer for 21 time steps, based on which the partitioning can be done. As input, the \ac{MLP} gets the descriptors of 11 layers, $1 R_e$ spaced apart. Three of them are in front of the current prediction layer. All the hyperparameters of the \ac{MLP} were optimized using the platform Weights and Biases. In total, the \ac{MLP} has around 65000 parameters.

\subsubsection*{Model assumptions and design rationale}
The central design choice is to formulate the prediction task on a per-layer basis rather than using the full domain as input. This is motivated by three main considerations: (i) independence of the model performance from the domain size in $z$-direction, (ii) enforced generalization across all layers, and (iii) a substantial increase in the number of training samples. The key assumption underlying this strategy is that the propagation of the structure-formation front -- the dominant source of divergence between simulation methods -- follows qualitatively similar dynamics throughout the simulation.

Compared to a global-domain architecture, the layer-wise formulation prevents the network from memorizing specific time–position patterns. Instead, it compels the \ac{MLP} to treat each layer consistently, irrespective of absolute position or simulation time. As a result, each occurrence of the structure-formation front is recognized and modeled in a similar manner.

This enforced invariance is particularly important for extending the decision model beyond the conditions present in the training set. Since training data must be generated on relatively small domains and over limited time horizons, the ability to generalize to longer domains and later time steps is essential. The layer-wise prediction not only removes the explicit dependence on the $z$-domain size, but also provides the necessary robustness for applying the model to simulations with larger spatial and temporal extents.

\subsubsection*{Production of training data}
\label{sec:training-data}
% subdomains of the simulation domain are relatively independent, so the discrepancy between training data and real simulation data is small enough

The application of the here-described \ac{ML}-model is to couple the full continuum simulations of the \ac{SNIPS} process to a locally confined particle simulation that overwrites the continuum model in a region where the latter is inaccurate. Hence, the goal of the \ac{ML}-model is to predict the error of the continuum simulation compared to the particle simulation, with the goal of determining the region in which it is least accurate to choose as a sub-domain for the particle simulation.

Optimal training data would mirror exactly the setting of the application. This would involve simulating a range of fully coupled simulations, with randomly partitioned domains, to include all sources of errors between the particle solution and the continuum solution. %\MM{No clear what you want to say with the 2 sentences above.} \MB{Just means that optimal training data is not feasible, hence we use approximations} 
This would introduce orders of magnitude of additional complexity and computational cost not only for prediction, but also for training data generation. Thus, we need to simplify the problem with the following assumptions:
\begin{itemize}
    \item The time evolution of a subdomain is relatively independent of its coupling. This means, for all of our scenarios, a subdomain of a particle simulation embedded into a continuum simulation behaves approximately like the same simulation being part of a full particle simulation. 
    \item The characteristics of the simulation are from a certain point onwards similar enough to be able to generalize to the later time steps. 
    \item Because of the reduction to descriptors, the \ac{MLP} is independent of the size of the domain in the lateral $x$ and $y$ directions (orthogonal to the movement of the structure-formation front). Hence, for the training data, one fixed lateral domain size is chosen. It is selected as small as possible, but such that all important characteristics are still captured.
\end{itemize}

\begin{figure}[tb!]
    \centering
    \includegraphics[width=\textwidth]{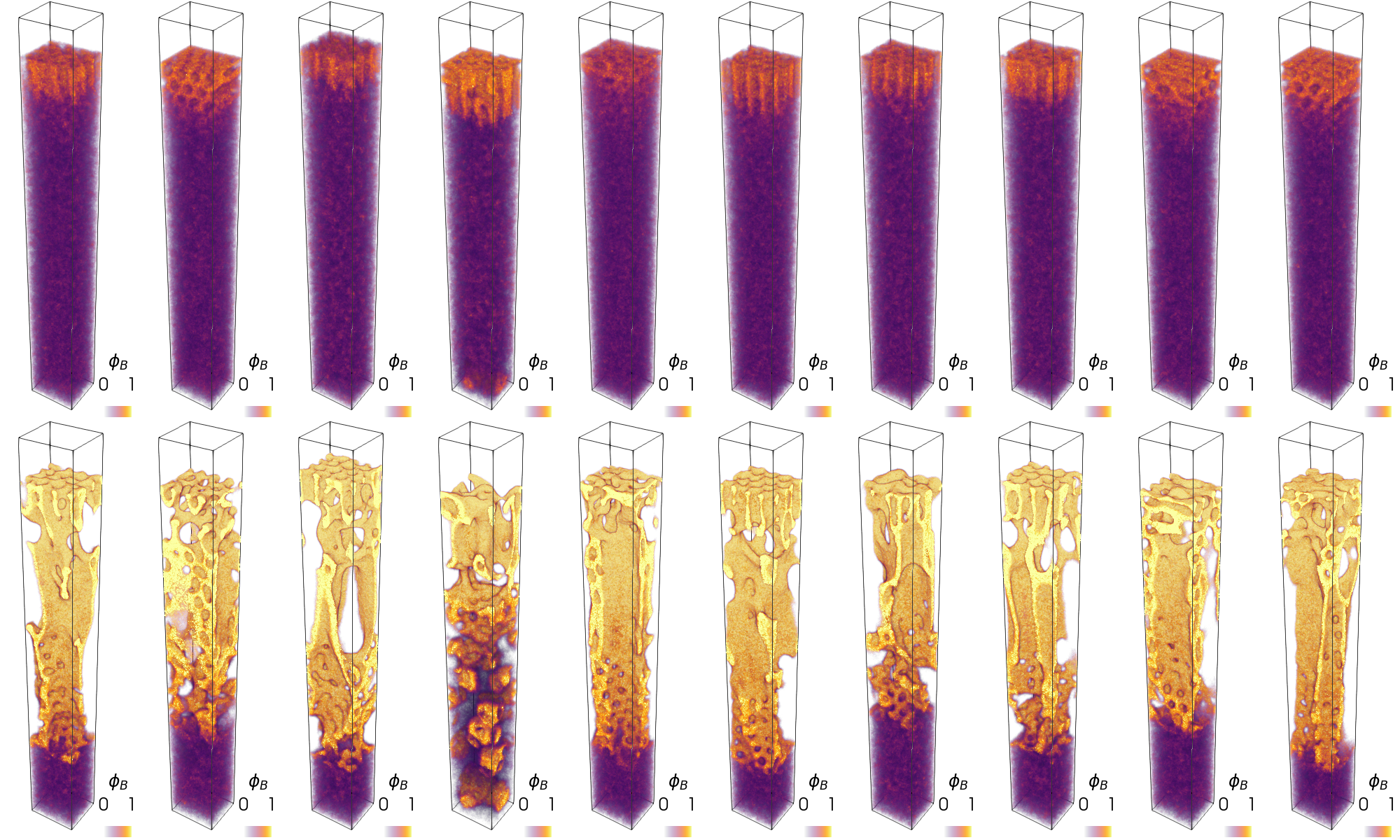}
    \caption{Example snapshots of 10 realizations (process parameters randomly varied between columns) of the training data, produced with the particle model. The top row shows the $B$-concentration fields, $\phi_B(\mathbf{r})$, after \ac{EISA}, the bottom rows shows these after \ac{NIPS}.}
    \label{fig:training-data}
\end{figure}
Our model is targeted to work on a large range of material and process parameters. For this reason, we generate 40 sets of parameter combinations that are close to the below base parameter combination but differ strongly enough that a number of different physically relevant morphologies emerge. To this end, we randomly generate the $\chi_{\alpha \beta}N_P$ interaction matrix from a Gaussian distribution, with the mean corresponding to the base-parameter combination and the standard deviation 
\begin{align}
    \sqrt{\langle \chi N_P^2\rangle - \langle \chi N_P \rangle ^ 2} = 
\begin{pmatrix}
  & A   & B   & S   & C   & G  & N \\
A & 0   & 2.5 & 2.5 & 0   & 15 & 5 \\
B & 2.5 & 0   & 0   & 1   & 25 & 25 \\
S & 2.5 & 0   & 0   & 0   & 10 & 10 \\
C & 0   & 1   & 0   & 0   & 10 & 10 \\
G & 15  & 25  & 10  & 10  & 0  & - \\
N & 5   & 25  & 10  & 10  & -  & 0
\end{pmatrix}
\end{align}
In particular, we do not vary the interactions $\chi_{AC}N_P$ and $\chi_{BS}N_P$, because this would allow positive values, \textit{i.e.}, repulsions of $A$ to $C$, and $B$ to $S$. In experimental systems, the majority block, $B$ typically has a preference for the volatile solvent $S$, while the minority block, $A$, typically is preferential towards the nonvolatile solvent $C$. All random parameter combinations should conform to this principle. Indeed, varying the above interaction, allowing positive values, resulted in many realizations for which the initial solution was immediately unstable to macrophase separation of polymer from the two solvents. This has no experimental relevance and, thus, also none to our simulation studies.

A particle simulation is run fully independently for each parameter combination, with $t_{\mathrm{EISA}} =  5\cdot 10^5\mathrm{MCS} \approx 16\tau_R$ and $t_{\mathrm{NIPS}}= 7.5\cdot 10^5\mathrm{MCS} \approx 24\tau_R$. The instantaneous concentration fields are saved every $t_{\mathrm{save}} = 5000\mathrm{MCS} \approx 0.16\tau_R$. Every second of these saved morphologies then serves as an initial condition to a continuum simulation that is run for $t=4.2\lambda_0^{-1}$, saving the concentration fields synchronously to the particle simulation, \textit{i.e.}, every $t_\mathrm{save} = 0.21\lambda_0^{-1}$. 

The largest bottleneck of the training data generation is the label generation since it requires a full particle simulation as a reference solution to compare the continuum simulation with. Thus, this method of training data generation is efficient, since only a single particle simulation is run per parameter combination, while the repeated start of the computationally efficient continuum simulation generates a large amount of comparable time evolutions between the two models. Afterwards, a sample of training data consists of 21 labels and a set of descriptors for multiple layers, like visualized in \autoref{fig:decision_model_architecture}. Finally, generating the labels for the 40 different parameter combinations results in a total of $1.2\cdot 10^6$ samples to train the \ac{MLP}.
% \GH{@Matthias: check if this last paragraph is still correct.} \MB{@Gregor: I made some changes, just structurally, are you ok with them?}

\subsection{Post-processing}
\label{postprocessing}
% explain the post-processing and show one exemplary result mean error plot
% one plot with multiple curves for different time steps in it/two plots for two simulations
\begin{figure}
    \centering
    \includegraphics[width=0.8\linewidth]{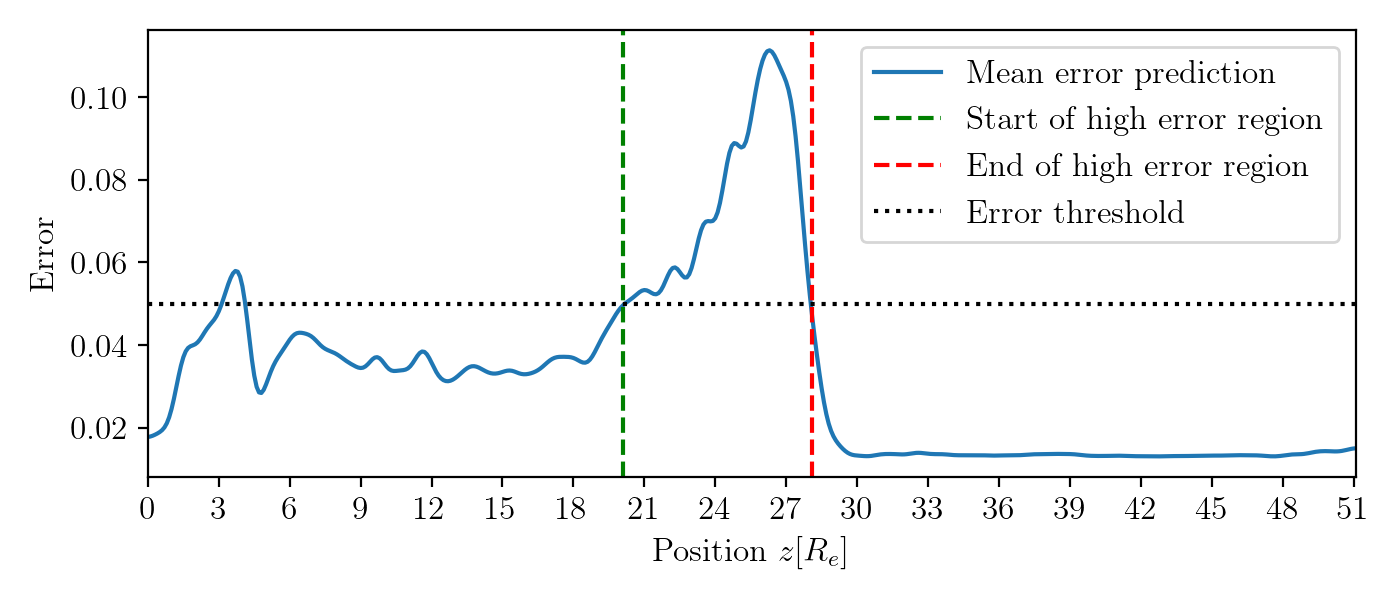}
    \caption{Mean error distribution for a single time step. The errors (\autoref{fig:error_plots}) are aggregated over a range of time steps approximately up to the point in time where the next prediction will be done. From this error distribution, one subdomain, where errors exceed the threshold the most, is identified as the part of the domain where the higher precision simulation is applied. The subdomain is marked by the green and red dashed lines, the threshold by a black dashed line.}
    \label{fig:siglemeanerror}% Simulation index 20, time step 25
\end{figure}
The goal of the post-processing step is to generate a reliable subdomain in which the higher-fidelity simulation must be applied. This subdomain is determined from the error predictions introduced in Section~\ref{prediction}. From the whole range of time steps available, a subset of approximately the last quarter until the next prediction update is chosen. This accounts for the relevance of errors until the next update as well as effectively reducing time constant errors by only incorporating time steps where error peaks have already formed. In practice, the procedure combines several operations: an adaptive thresholding of the predicted error, identification of the dominant connected error region, Gaussian smoothing to reduce local noise, and aggregation of errors across the considered time horizon.

An exemplary result is shown in \autoref{fig:siglemeanerror}. Here, the ordinate represents the mean error, aggregated over multiple prediction steps, and the abscissa corresponds to the $z$-position of the layer. The black dashed line indicates an exemplary threshold at a value of $0.05$. The resulting subdomain is bounded by green and red dashed lines, marking the beginning and end of the high-error region, respectively. Only one connected region is selected, namely the one with the highest error peak. In the illustrated case, two peaks are present in the mean error distribution. The higher peak is enclosed by the marking lines and corresponds to the structure-formation front, averaged over multiple time steps. In contrast, the second peak to the left is associated with the majorly solidified solution top layer. In the time evolution of \autoref{fig:decision_model_architecture} it becomes visible that this peak appears immediately at the beginning of the continuum simulation and afterwards retains the same height. This means, there are some immediate short-time dynamics upon initialization with particle-simulation concentration fields, but no long-time changes occur, not requiring a long-time outsourcing of this region to the particle simulation. To the right of the structure-formation front, the mean error is close to zero, reflecting the homogeneous fluid.

\section{Results}
\subsection{Error prediction results}
\label{sec:error-prediction-results}

\begin{figure}[t!]
    \centering
    \begin{subfigure}{0.49\textwidth}
        \includegraphics[width=\linewidth]{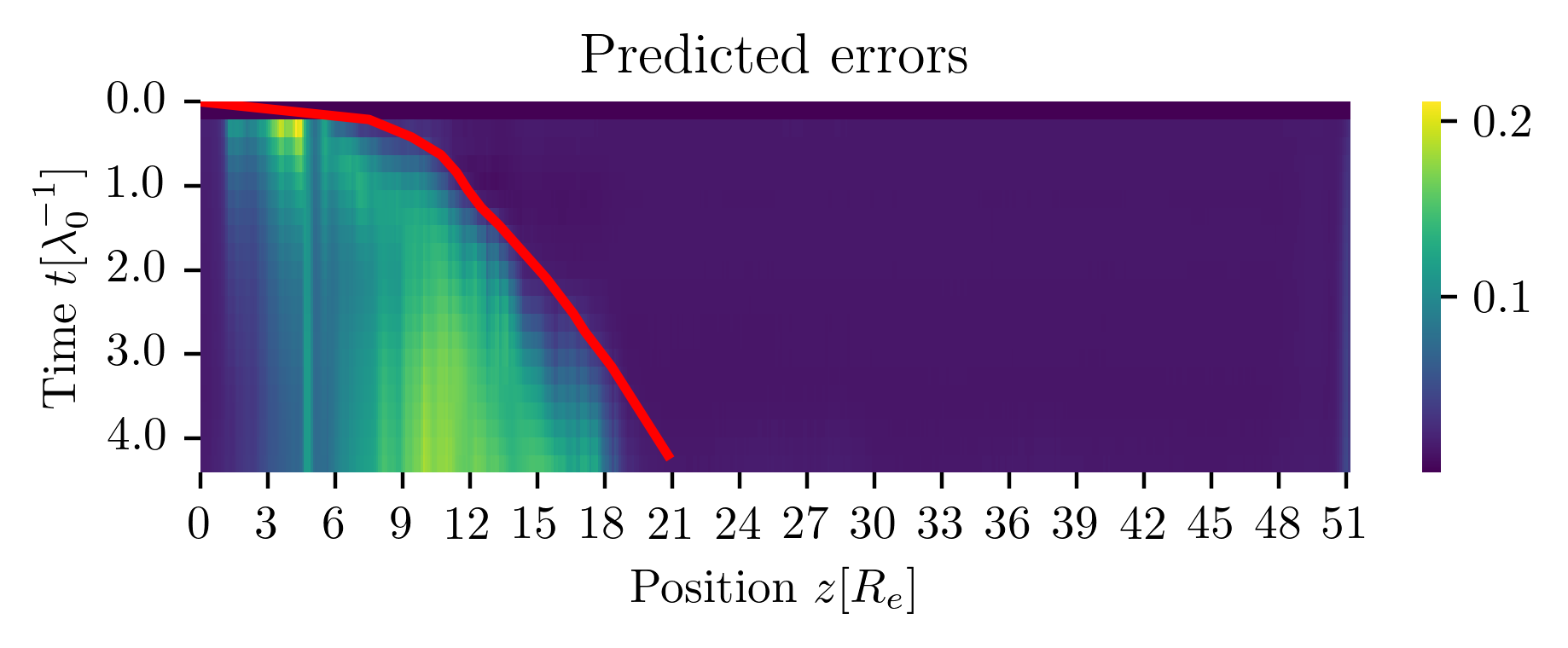}
    \end{subfigure}
    \hfill
    \begin{subfigure}{0.49\textwidth}
        \includegraphics[width=\linewidth]{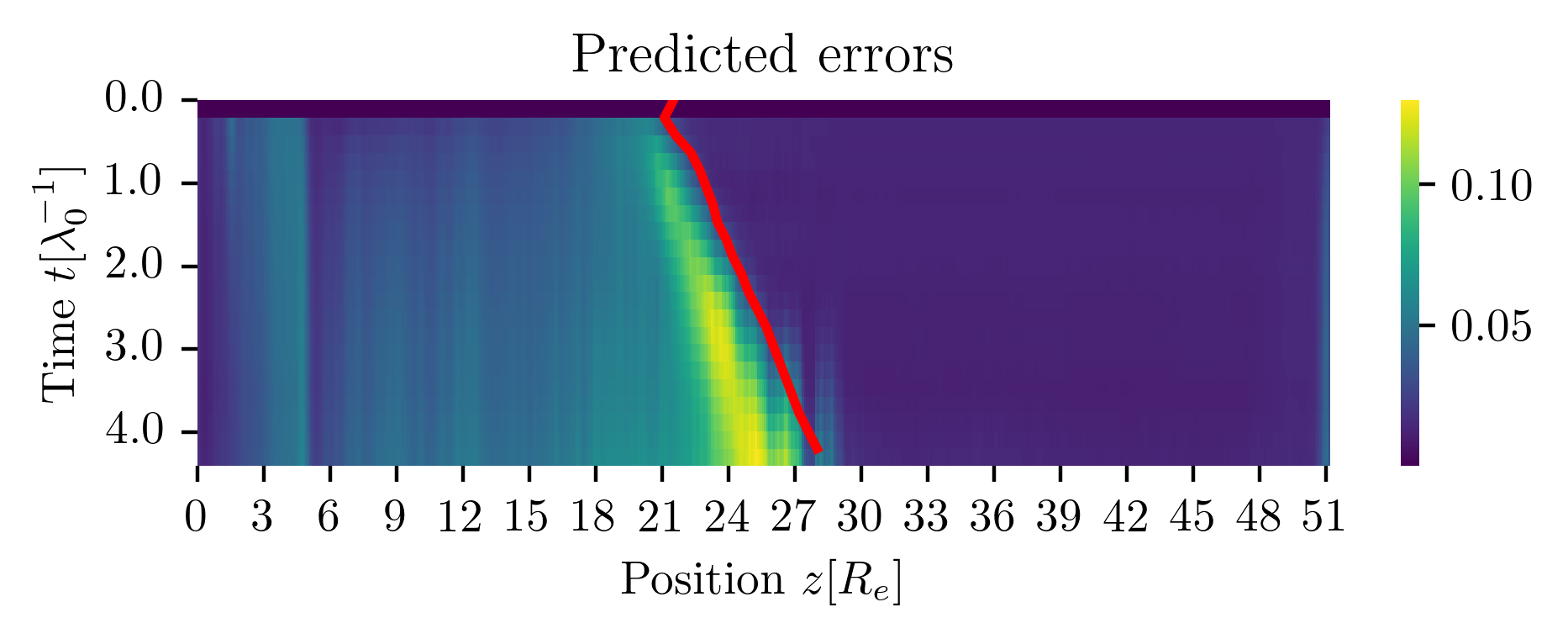}
    \end{subfigure}
    \vspace{0.5cm}
    \begin{subfigure}{0.49\textwidth}
        \includegraphics[width=\linewidth]{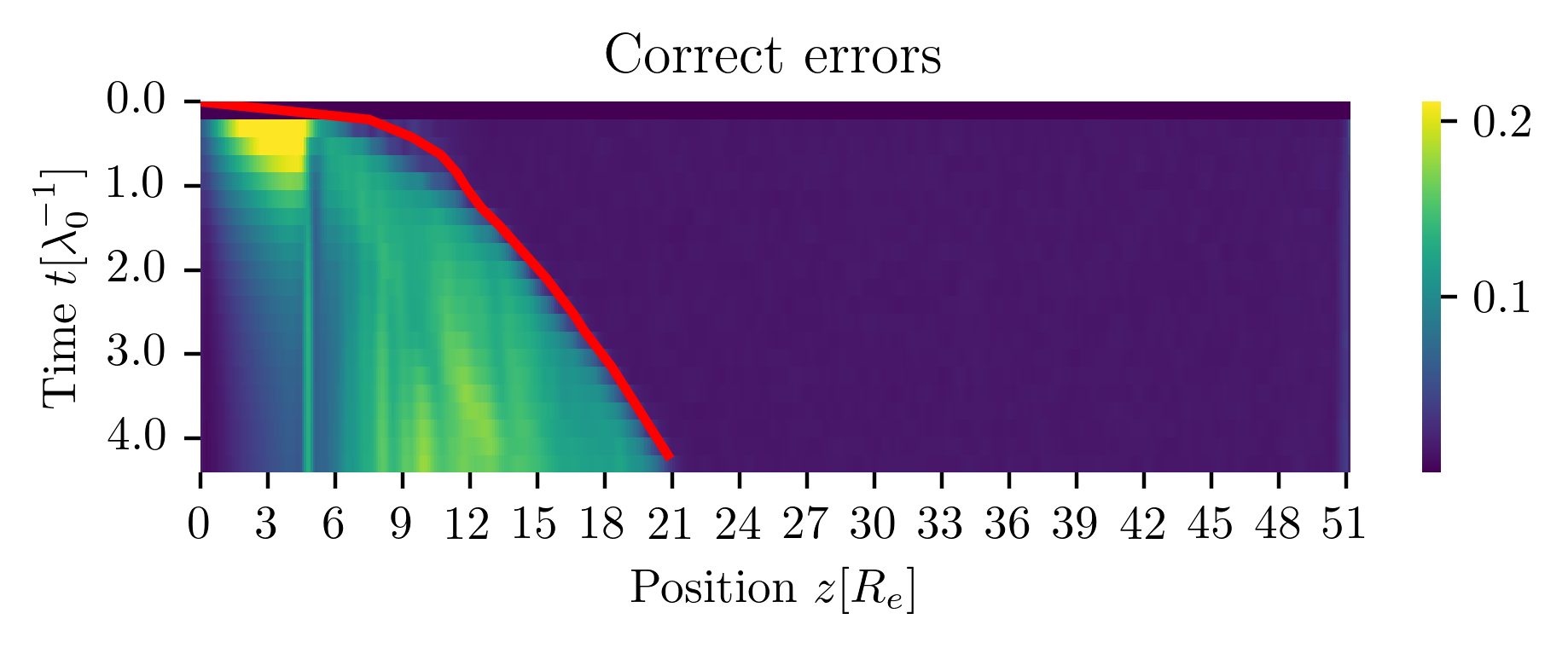}
        \caption{NIPS process beginning at time $T=0\lambda_0^{-1}$}
    \end{subfigure}
    \hfill
    \begin{subfigure}{0.49\textwidth}
        \includegraphics[width=\linewidth]{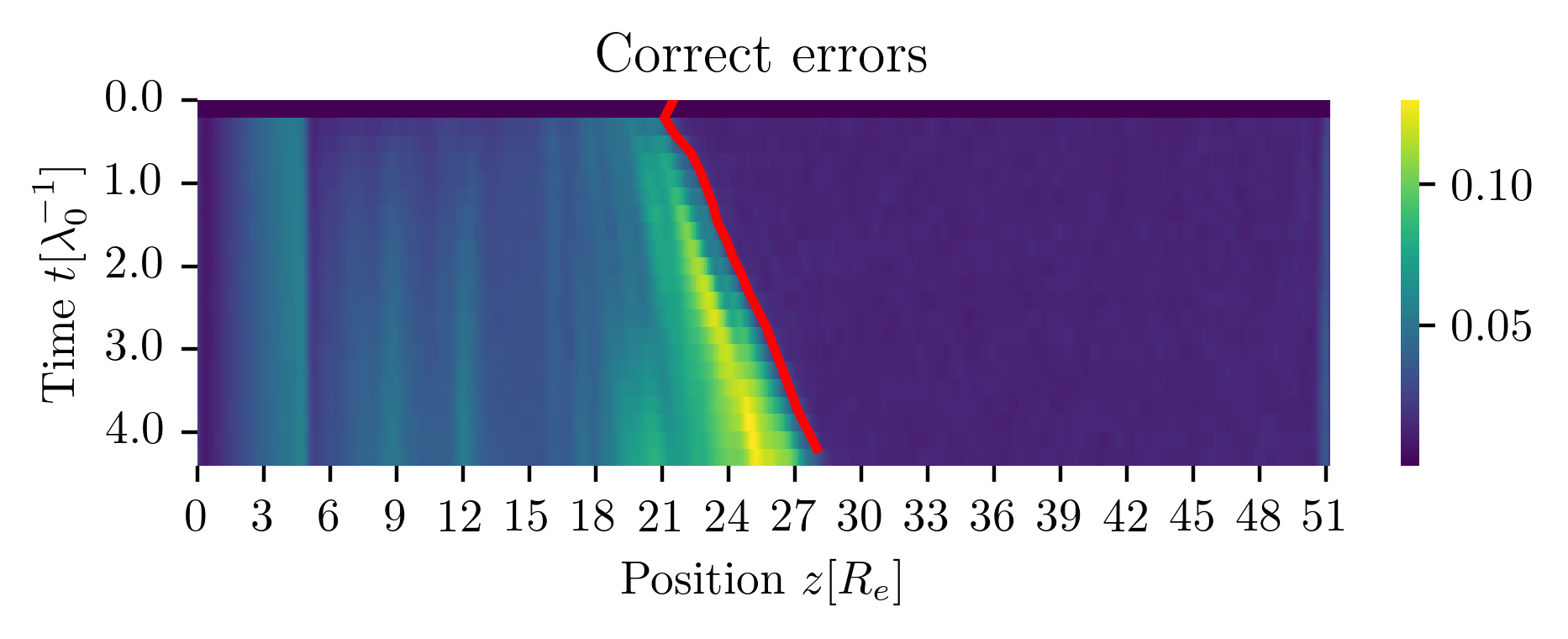}
        \caption{NIPS process beginning at time $T=4.2\lambda_0^{-1}$}
    \end{subfigure}
    \begin{subfigure}{0.49\textwidth}
        \includegraphics[width=\linewidth]{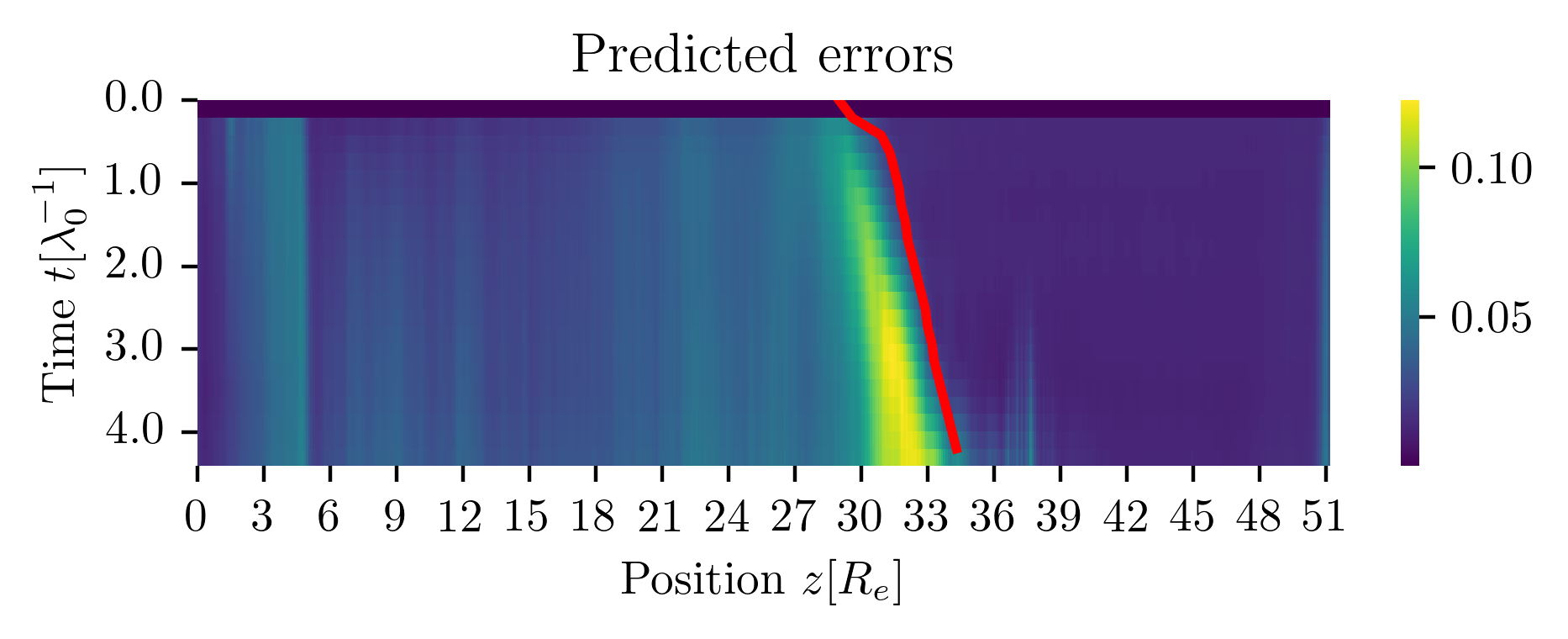}
    \end{subfigure}
    \hfill
    \begin{subfigure}{0.49\textwidth}
        \includegraphics[width=\linewidth]{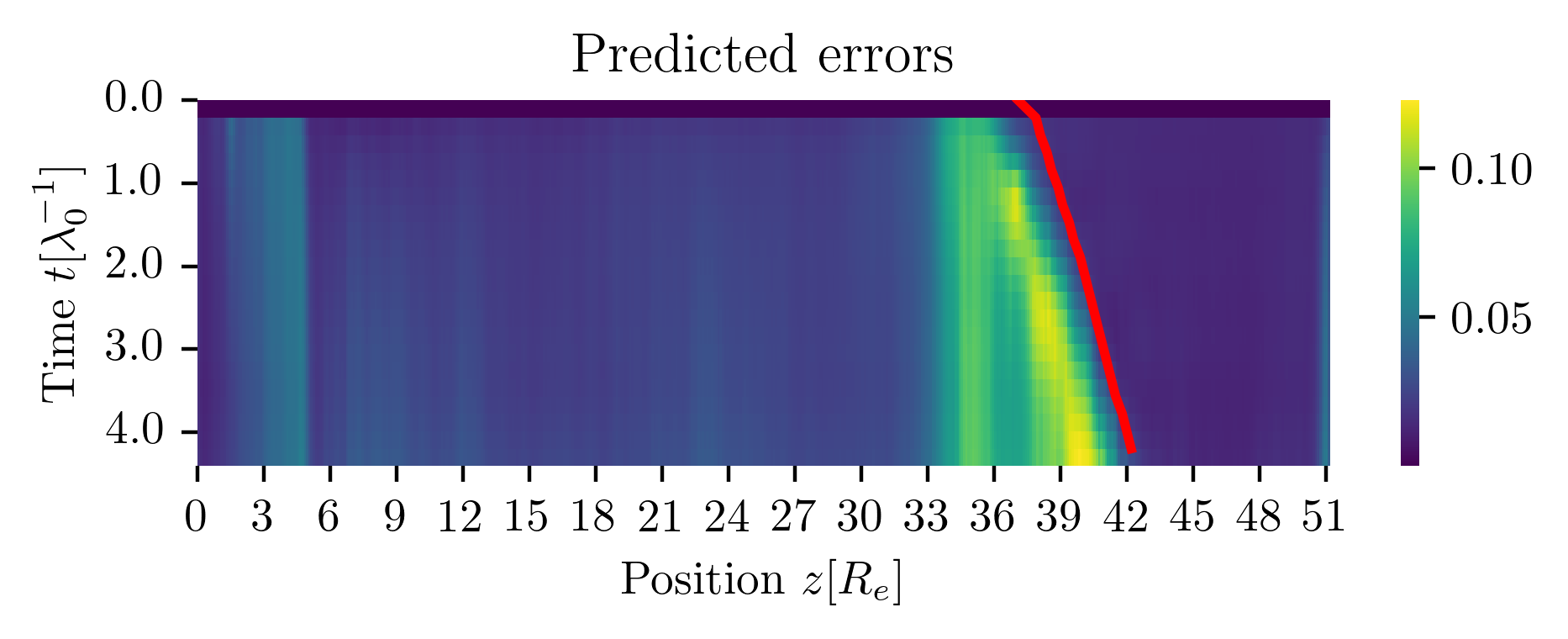}
    \end{subfigure}
    \vspace{0.5cm}
    \begin{subfigure}{0.49\textwidth}
        \includegraphics[width=\linewidth]{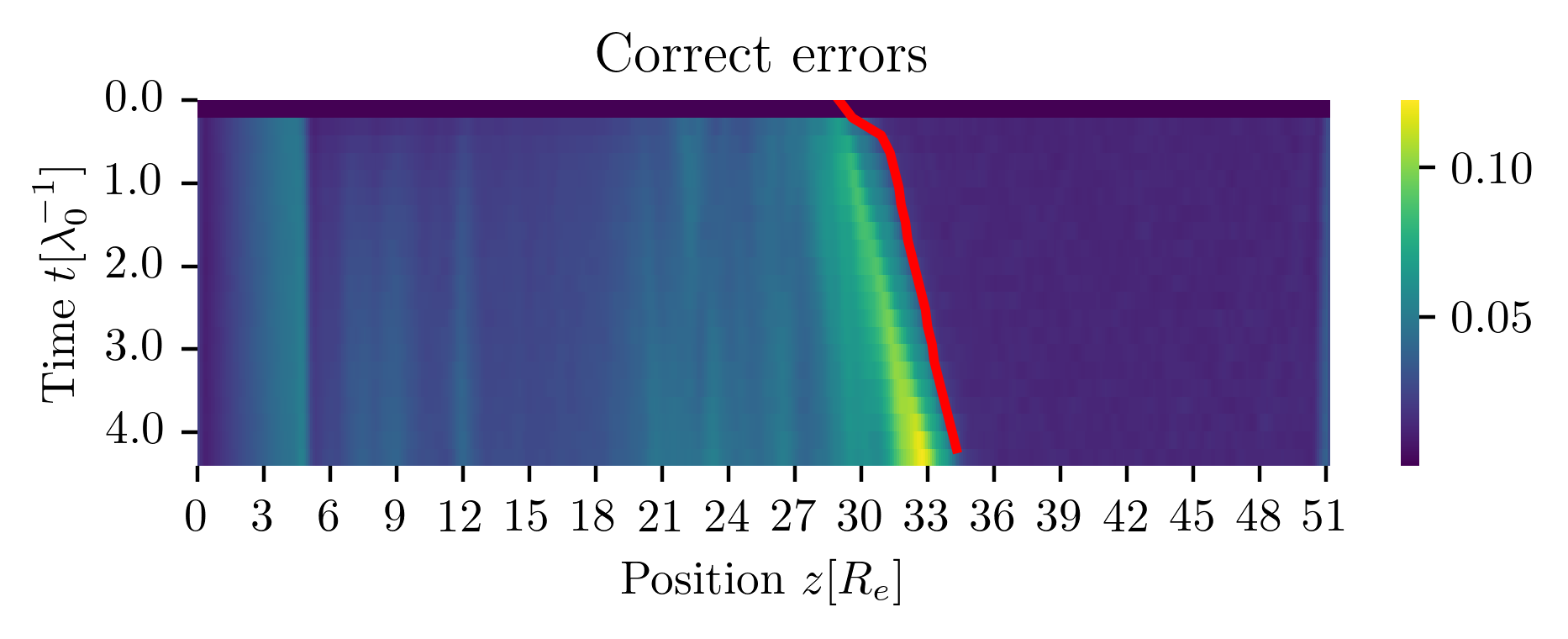}
        \caption{NIPS process beginning at time $T=8.4\lambda_0^{-1}$}
    \end{subfigure}
    \hfill
    \begin{subfigure}{0.49\textwidth}
        \includegraphics[width=\linewidth]{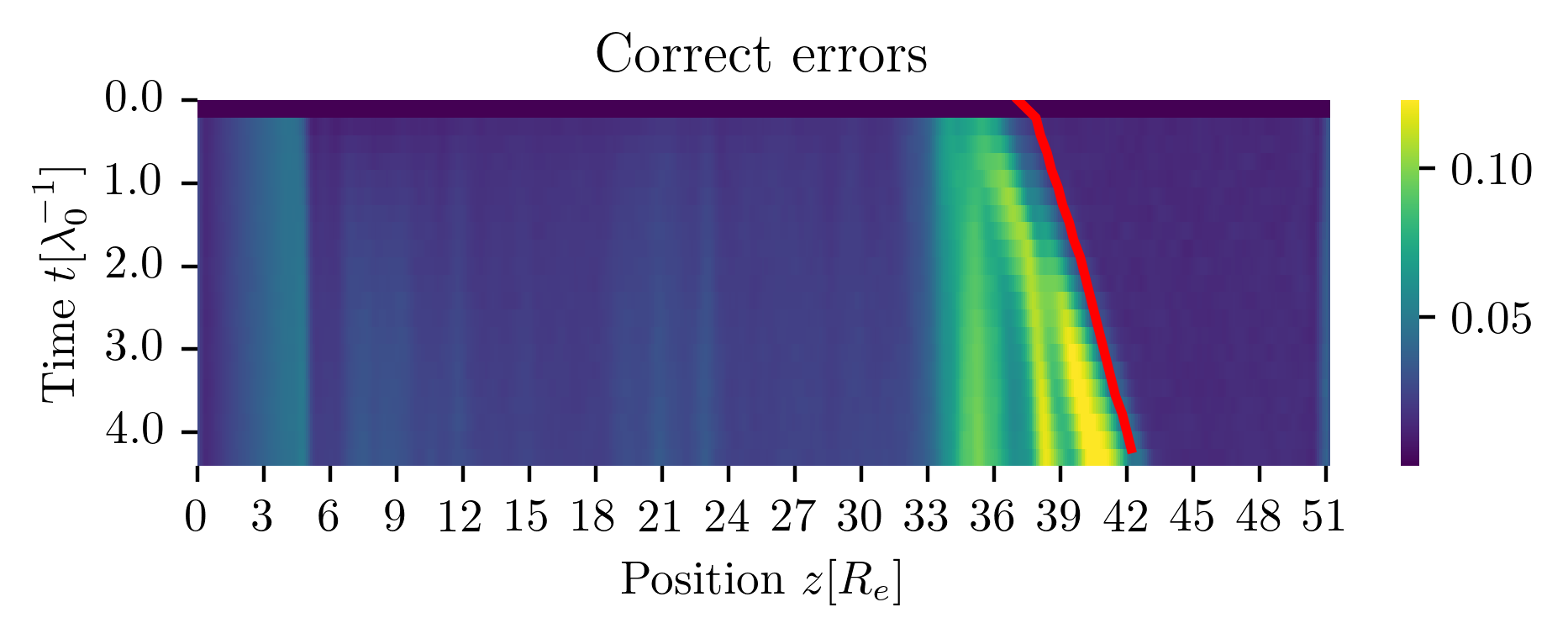}
        \caption{NIPS process beginning at time $T=12.6\lambda_0^{-1}$}
    \end{subfigure}
    \caption{Heatmaps with predicted errors (upper images) in comparison to the correct errors (lower images). In each heatmap, the red line indicates the position of the structure-formation front, estimated by using the non-solvent density threshold $\phi_N^\ast=0.0017$, at which instability against macrophase separation emerges (see \autoref{fig:triangle} in \ref{sec:gibbs-triangle-testing-data} for the full Gibbs triangle). %\JX{1. Just to confirm: is this (0.0017) the non-solvent density threshold value used for plotting the red lines? 2. How would we like to implement the cross-reference between main and SI?} 
    The data comes from an exemplary simulation which is not in the training set. For other simulations, predicted and correct errors exhibit similar dynamics. On the $x$-axis the layer coordinate shown, meaning this is the direction the structure-formation front is moving. The $y$-axis denotes the relative time step to the starting configuration, which is noted below the respective diagrams. }
    \label{fig:error_plots} % simulation index 19
\end{figure}

In \autoref{fig:error_plots} a comparison between calculated and predicted errors is shown for a set of different time points during the NIPS simulation. The data stems from a simulation in the test set and was therefore not included in the \ac{MLP} training. On the abscissa, the $z$-position (layer coordinate) is shown, while the ordinate denotes the time relative to the current simulation state as indicated in the respective captions. The red line marks an approximation of the structure-formation front, estimated by using the non-solvent density threshold $\phi_N=0.0017$, at which instability against macrophase separation emerges on the Gibbs triangle as predicted by RPA (see \autoref{fig:triangle}). It shows that generally, the error between both simulations begins to grow directly where the structure formation starts. 

The predicted errors show overall good qualitative agreement with the reference errors. In particular, the movement of the structure-formation front is clearly captured by the \ac{MLP}, with error peaks shifting accordingly in space and time. The model reproduces the important features such as the location of maximum errors, the shape of error peaks, and the extent of high-error regions. Similarly, low-error regions are predicted with good accuracy.  

Deviations appear mainly at later prediction times, which is expected, since forecasting further into the future generally becomes more challenging. At the beginning of the simulation, the very first predictions (\autoref{fig:error_plots} a) also show notable discrepancies, likely because the simulation characteristics are not yet well established at this early stage. Nevertheless, even here the qualitative shape of the error distribution is reproduced. Finally, at the latest prediction time steps, some artifacts emerge in front of the structure formation region, located just ahead of the high-error zones (mainly visible in \autoref{fig:error_plots} b and c). While these artifacts are not present in the ground truth, they remain confined to the later stages of the prediction. They might have their origin in noise of the input data and an unclear estimation of the structure-formation fronts' propagation speed.

\subsection{Post-processing results}
\begin{figure}[t!]
    \centering
    \begin{subfigure}{0.49\textwidth}
        \includegraphics[width=\linewidth]{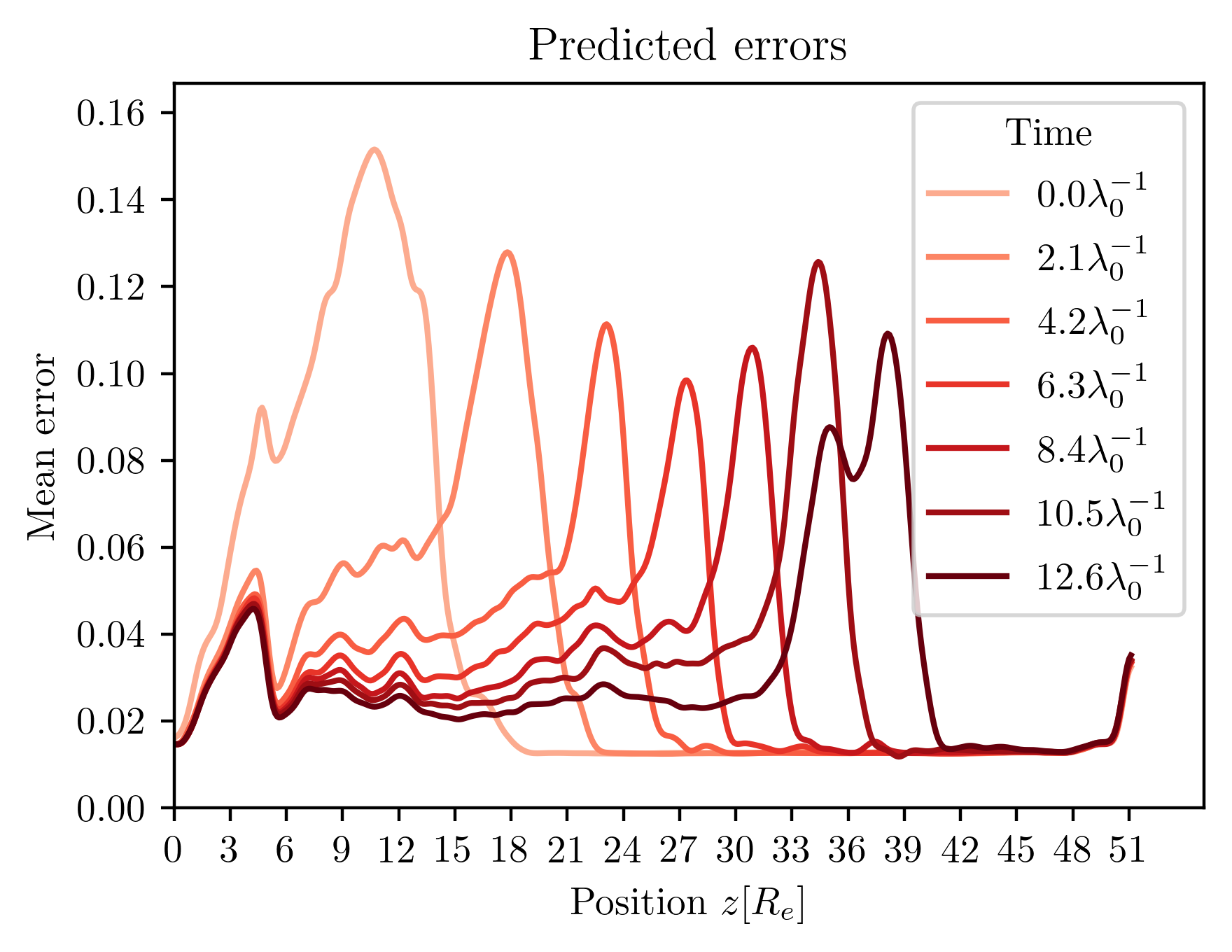}
    \end{subfigure}
    \hfill
    \begin{subfigure}{0.49\textwidth}
        \includegraphics[width=\linewidth]{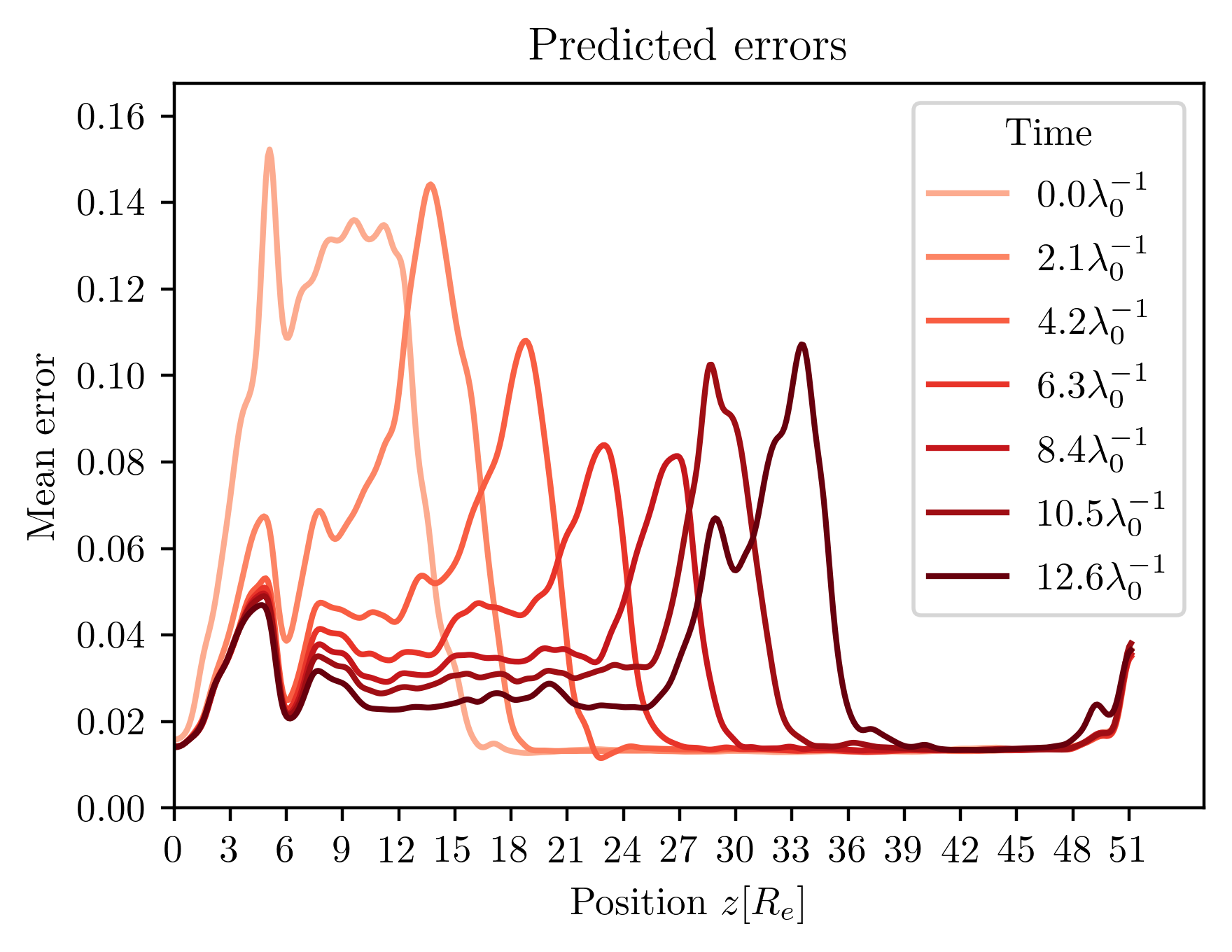}
    \end{subfigure}
    \vspace{0.5cm}
    \begin{subfigure}{0.49\textwidth}
        \includegraphics[width=\linewidth]{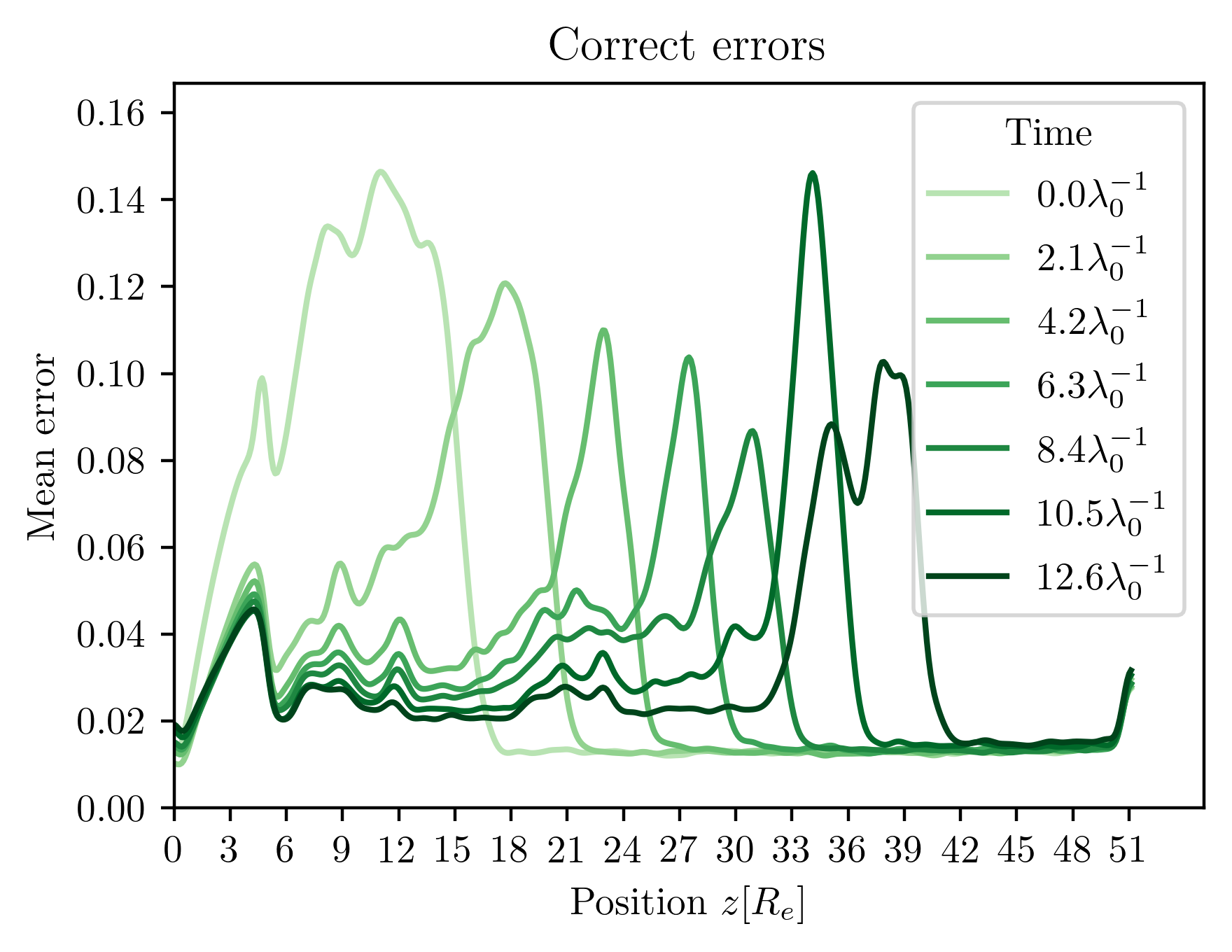}
    \end{subfigure}
    \hfill
    \begin{subfigure}{0.49\textwidth}
        \includegraphics[width=\linewidth]{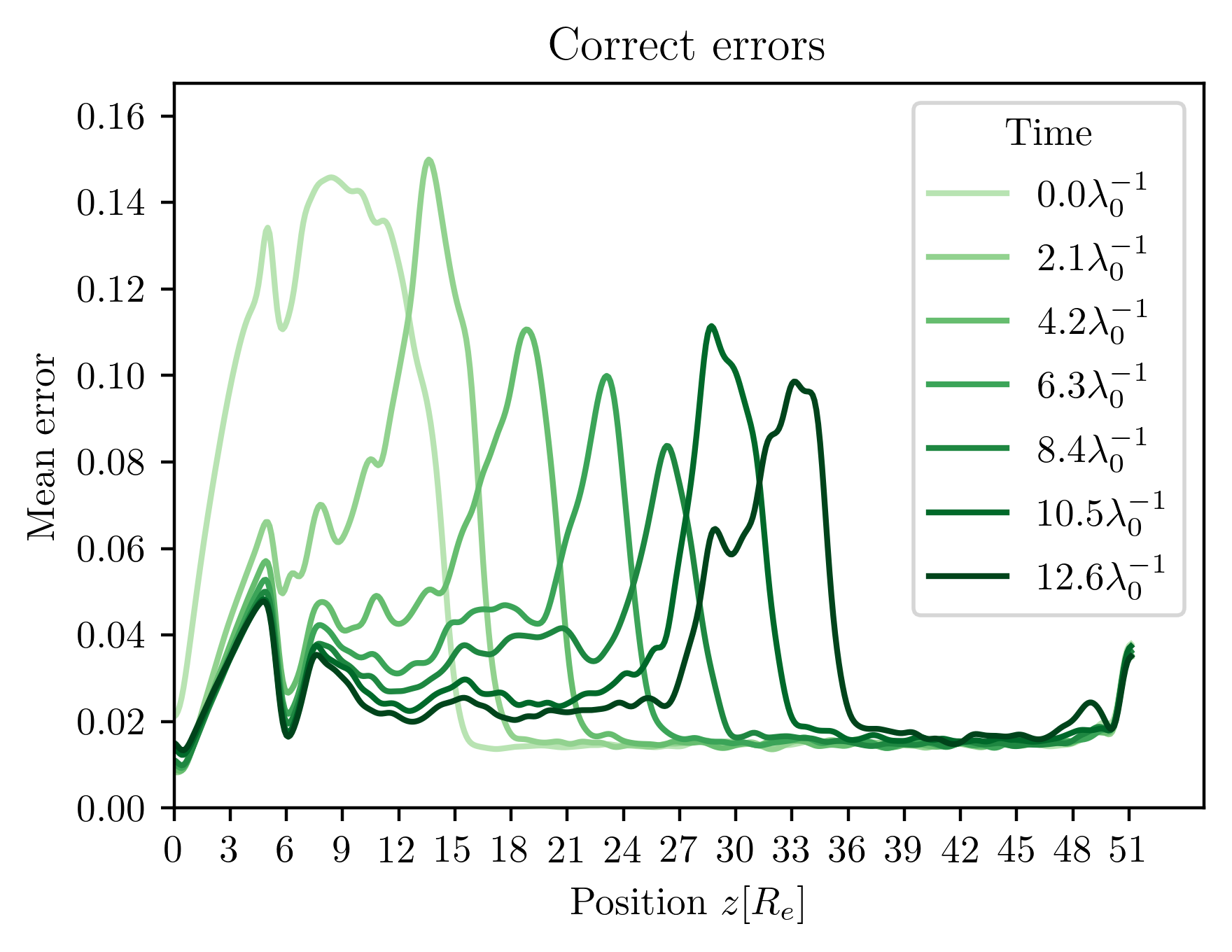}
    \end{subfigure}
    \caption{Mean error distributions for different time steps of two simulations. The respective upper diagram depicts the predictions (red colors), the lower diagram depicts the correct error values (green colors). Between the left and right diagrams the simulation parameterization is changed. Both simulations are excluded from the training data. }
    \label{fig:meanerrors}% Simulation index 19 (left) and 18 (right)
\end{figure}

A more comprehensive comparison between predicted and reference post-processed error distributions is given in \autoref{fig:meanerrors}. Results are shown for two test simulations not included in the training data. In each case, the upper diagrams display the predicted error distributions at several time steps, while the lower diagrams show the corresponding ground-truth errors. Within each plot, the temporal evolution is illustrated by curves of varying color intensity. The comparison demonstrates that the predicted mean error distributions reproduce the key qualitative features of the reference: the location of the structure-formation front, the movement of the error peaks over time, and the characteristic shape of the distributions. Both simulations exhibit different speeds of front propagation as well as different magnitudes and temporal changes of the error. These differences are captured with good fidelity by the predictions, underlining the robustness of the post-processing approach.

\subsection{Shapley additive explanations (SHAP) analysis}
\begin{figure}[tb!]
    \centering
    \includegraphics[width=0.8\linewidth]{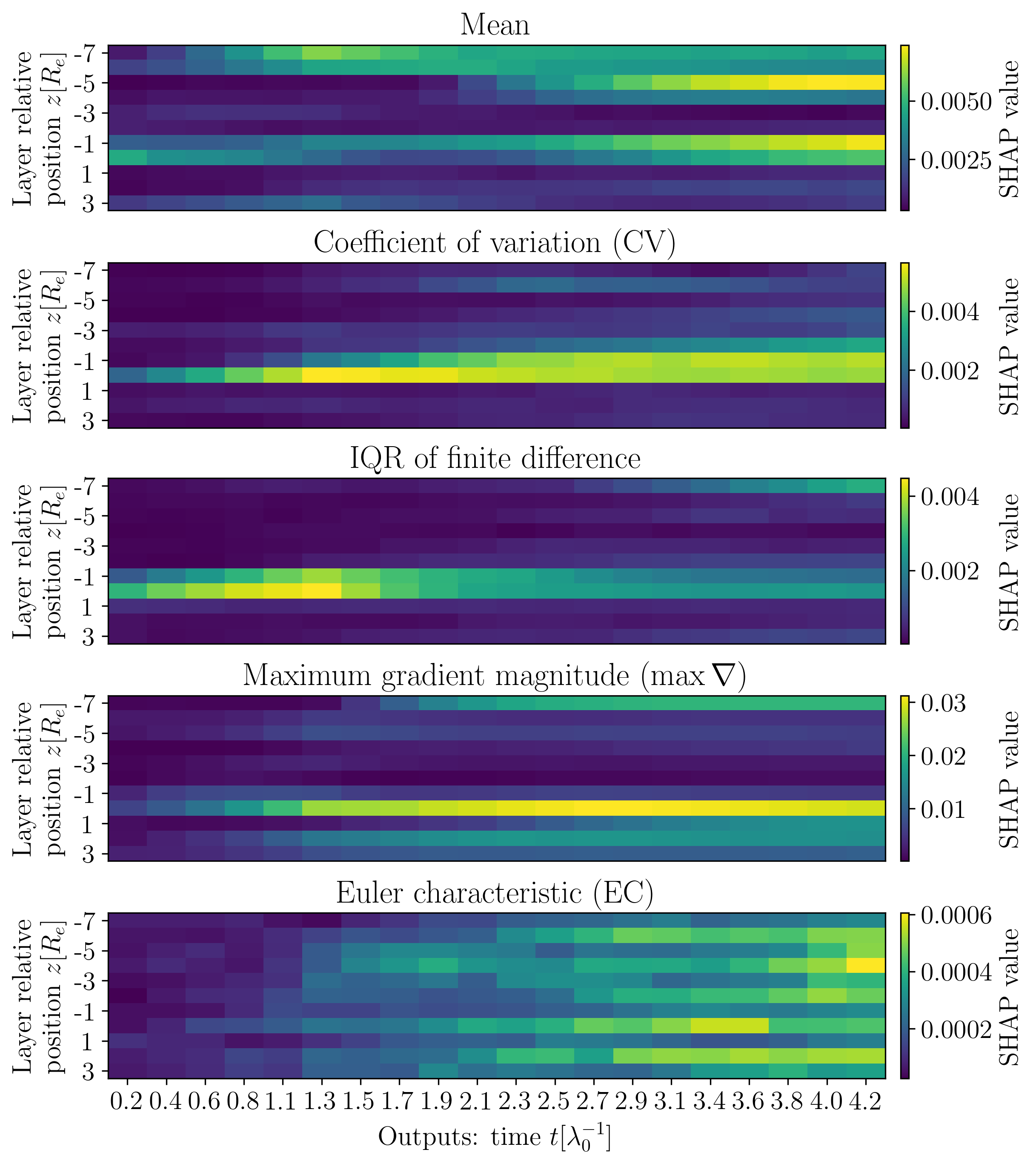}
    \caption{Results of the Shapley additive explanations (SHAP) analysis, sorted separately for each descriptor. Notably, the scale of the values varies significantly as can be seen at the color bars. All diagrams share the same $x$-axis, describing the different \ac{MLP} outputs whereas the $y$-axis denotes the positions of the different input layers, relative to the current layer the prediction is made for. }
    \label{fig:shap}
\end{figure}

To better understand how the prediction \ac{MLP} uses the input descriptors, we applied Shapley additive explanations (SHAP) to quantify the contribution of each input feature to each predicted output. The resulting SHAP matrix (\autoref{fig:shap}) contains the mean contribution values across all samples, sorted by descriptor. Each descriptor is shown with its own color scale, as absolute SHAP magnitudes vary by up to two orders of magnitude. The $y$-axis represents the spatial offset of the input layers relative to the target layer in units of $R_e$, with $0Re$ denoting the layer for which the divergence is predicted. The $x$-axis indicates the relative prediction time step. This representation allows us to identify both which descriptors are most influential overall and how their relevance changes across space and prediction horizon.

The SHAP analysis results show clear trends in how the MLP utilizes the different descriptors. As expected, the layer corresponding to the target position ($0R_e$) is generally the most influential, indicating that information from the current layer is most relevant for predicting its own divergence. For most descriptors, the SHAP values are small for short prediction horizons and increase for later output times, which is consistent with the observation that prediction errors also grow with the forecasting horizon.

Among the descriptors, the maximum gradient magnitude stands out as the most important by a substantial margin, with SHAP values roughly one order of magnitude larger than those of other descriptors. This descriptor is also the only one for which the MLP makes use of information from as far as three layers ahead of the target layer -- likely reflecting its strong predictive value before the onset of structure formation. At later time steps, this descriptor is also used for the last available layer, which may relate to its relevance for estimating large errors far into the future. A similar pattern is observed for the interquartile range (IQR) of finite differences, which is most heavily used at the current and previous layer for early prediction times, suggesting that local structure formation directly informs near-term error growth.

The coefficient of variation (CV) shows a similar spatial usage pattern to the IQR, being most relevant at the current and previous layers, but it also retains moderate importance for longer-term predictions. The mean descriptor, in contrast, is primarily used for the most distant prediction horizons, indicating that large-scale average values play a role when detailed local information becomes less predictive.

Finally, the Euler characteristic shows consistently negligible SHAP values across all outputs and layers. Its contribution appears to be close to noise level, which may indicate that it is not a reliable predictor in this context or that any apparent signal is the result of overfitting in the descriptor selection process.

%The SHAP analysis (Fig. \ref{fig:shap}) shows that the predictive \ac{MLP} relies most strongly on information from the target layer ($0R_e$), with feature importance generally increasing for longer prediction horizons. Among all descriptors, the maximum gradient magnitude is by far the most influential, with SHAP values about an order of magnitude larger than those of other inputs. Several descriptors, including the coefficient of variation and the interquartile range of finite differences, are primarily used at the current and preceding layers, while the mean becomes more relevant for the most distant predictions. The Euler characteristic contributes negligibly in all cases. A more detailed interpretation is provided in \autoref{Asec:shap}.

\subsection{Inference to continuum-simulation results}

The training of the \ac{ML}-model is based on particle simulations and the dynamics, from arbitrary transient morphologies as initial conditions, within the continuum model. Now, let us apply the \ac{ML}-based error prediction to the continuum model simulation presented above. Here we choose the one with weaker solvent attraction, $\chi_{SC}N_P=-28$, compared to the base parameters, \autoref{fig:improved-parameters-morphology}, which resulted in a continuous membrane morphology in both simulation models. 

\begin{figure}
    \centering
    \includegraphics[width=\textwidth]{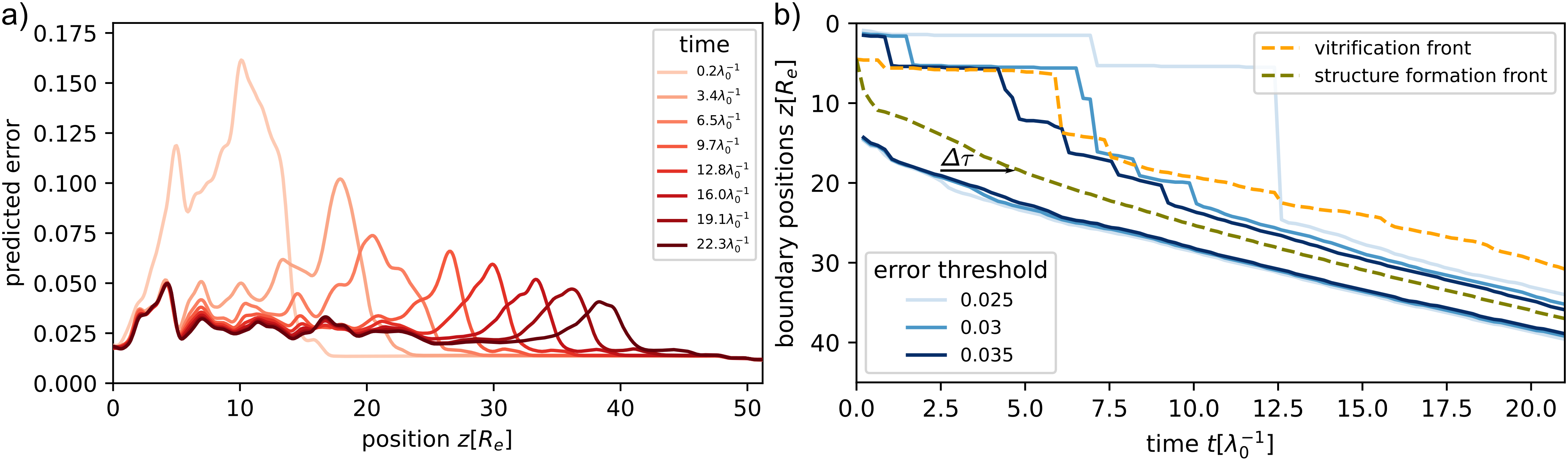}
    \caption{a) Error prediction of the continuum simulation compared to the particle simulations based on the pure-continuum simulation of \autoref{fig:improved-parameters-morphology}. The figure shows the mean error at different times during the \ac{NIPS} process, plotted against the vertical position, $z$. b) Resulting sub-domain selection after postprocessing for different threshold values (solid) and for the region with the persistent peak in the cylindrical top-layer (around $z=4R_e$, dashed).}
    \label{fig:app-to-udm}
\end{figure}
\autoref{fig:app-to-udm} a) shows the mean error prediction for the error after time $\Delta \tau = 2.4\lambda_0^{-1}$, for the continuum simulation at different times during the \ac{NIPS} process. Generally, the error prediction for the full continuum simulation adopts the same shape as the prediction for the test data, \autoref{fig:meanerrors}, also in the case of a larger time horizon compared to the training data. The largest error is predicted to occur at early times at the top of the solution film. The error peak then travels through the film, but a minor peak remains at the top of the film. 

\autoref{fig:app-to-udm} b) shows the position of the proposed sub-domains for three different error thresholds, based on the error prediction. As a comparison, the figure shows the approximation of the vitrification and structure-formation front, introduced in \autoref{sec:membrane_fabrication}. As indicated, the \ac{ML}-proposed boundaries bracket the structure-formation front and lead by exactly the distance that the front travels during the outsource time $\Delta \tau$. This means that the \ac{ML}-model accurately predicts the speed of the structure-formation front. Furthermore, this simulation exceeds the time horizon of the training and shows that the error prediction generalizes to larger time steps, which are not included in the training data.

On the other end of the subdomain, the vitrification front is encapsulated at the beginning of the simulation, whereas at the later times it is excluded. Hence, the \ac{ML}-model predicts the later stages of structure formation, just before vitrification occurs with sufficient precision in the UDM.

\section{Discussion}

This study introduced a hybrid simulation scenario of a membrane fabrication process combining two complementary models: a particle approach providing high fidelity at the cost of computational speed, and a continuum-based approach offering higher efficiency at reduced accuracy. The investigated simulation scenario is characterized by a structure-formation front propagating along one coordinate direction, which serves as a representative test case for adaptive multi-model coupling.

To enable dynamic switching between these models, an error prediction mechanism based on machine learning was developed. As shown in \autoref{fig:decision_model_architecture}, the proposed decision model leverages two-dimensional descriptors and a layer-wise prediction strategy to estimate local model errors. These predictions form the basis for adaptive model selection, ensuring that computational resources are concentrated where accuracy is most critical while maintaining overall efficiency.

The results of the prototype implementation confirm that the proposed decision model enables adaptive coupling of two different simulation models in practice. As illustrated in \autoref{fig:decision_model_architecture}, the architecture bases decisions on predicted errors and thereby achieves efficient resource allocation while maintaining fidelity in regions of interest.

A necessary ability of the model is to generalize to a wide range of process- and material-parameter combinations of \ac{NIPS}, which the proposed architecture satisfies. It performs independently of the domain size in all three coordinate directions, extrapolates to later time steps not included in training, and remains robust under varying simulation parameters within the scope of the present study. These properties distinguish the approach from more narrowly parameterized surrogate models.

The error predicting \ac{MLP} and the layer-wise prediction architecture, however, are closely tied to the characteristics of the underlying system. In our case, the presence of a structure-formation front with localized dynamics motivated the choice of simplification to a 1D error prediction and localization strategy. Applications to other systems will require adapting these components to the relevant physical features.

Overall, the results demonstrate that machine-learned error prediction provides a viable route toward scalable, adaptive multi-fidelity coupling. Future work will integrate the present prototype in fully coupled simulations and systematic benchmarks, with the aim of establishing general guidelines for applying the framework to other classes of multiscale problems.

% \begin{itemize}
%     \item Advantages of \ac{ML}-model: Direct prediction of structure-formation front + size of possible sub-domain based on error + time evolution of these.
%     \item Physical interpretation of descriptors here? 
% \end{itemize}

\section{Acknowledgements}
We gratefully acknowledge the support of the Helmholtz-Gemeinschaft Deutscher Forschungszentren (HGF). The authors gratefully acknowledge the Gauss
Centre for Supercomputing e.V. (www.gauss-centre.eu) for providing computing time through the John von Neumann Institute for Computing (NIC) on the GCS Supercomputer JUWELS at J\"ulich Supercomputing Centre (JSC).

\section{Funding}
Financial support was provided by the Bundesministerium für Forschung, Technologie und Raumfahrt (BMFTR) within the project 16ME0658K \linebreak MExMeMo. 

\section{Declaration of competing interests}
The authors declare no competing interests. 

\section{Data availability}\label{dataavail}
The training data for the machine learning model, the model itself, as well as workflows and input files for the two simulation schemes and training data generation are provided on Zenodo: \url{https://doi.org/10.5281/zenodo.17396869}. 
%TODO: The data is available in the repository . 

\section{Code availability}\label{codeavail}
The source code for reproducing the results in this publication is provided on Zenodo: \url{https://doi.org/10.5281/zenodo.17396869}. 
%TODO: The code is available in the repository \url{}. 

\section{Reproducibility statement}% laden wir auch alles von den simulationen hoch oder nur die ML-sachen?
With the provided code and data, the results are reproducible.
\autoref{fig:descriptors} and % simulation index 20 
\autoref{fig:siglemeanerror} use data from the simulation with 0-based index 20, % Simulation index 20, time step 25
\autoref{fig:error_plots} from simulation 19, and % simulation index 19
\autoref{fig:meanerrors} % Simulation index 19 (left) and 18 (right)
from simulations 19 (left) and 18 (right).

\section{Author contribution: CRediT}
\textbf{Matthias Busch}: Conceptualization, Methodology, Software, Investigation, Formal Analysis, Writing - Original Draft, Visualization.
\textbf{Gregor Häfner}: Conceptualization, Methodology, Software, Investigation, Formal Analysis, Writing - Original Draft, Visualization.
\textbf{Jiayu Xie}: Methodology, Writing - Review \& Editing.
\textbf{Marius Tacke}: Writing - Review \& Editing.
\textbf{Marcus Müller}: Funding Acquisition, Supervision, Conceptualization, Writing - Review \& Editing.
\textbf{Christian J. Cyron}: Funding Acquisition, Writing - Review \& Editing.
\textbf{Roland C. Aydin}: Supervision, Conceptualization, Writing - Review \& Editing.

\section{Declaration of generative AI and AI-assisted technologies in the writing process}
During the preparation of this work, the authors used ChatGPT in order to review the manuscript for errors and possible improvements. This includes rephrasing parts of the manuscript. After using this tool/service, the authors reviewed and edited the content as needed and take full responsibility for the content of the published article.

\appendix

\section{Simulation techniques}
\label{sec:simulation-techniques-si}
\subsection{Continuum Model}
\label{sec:continuum_model-si}

The continuum model employs a free-energy functional approach based on the Uneyama-Doi free-energy functional that generalizes the Ohta-Kawasaki theory for diblock copolymers \cite{ohta_equilibrium_1986} to the strong segregation regime and blends with homopolymers and solvents. The system is described by normalized local concentration fields $\phi_\alpha(\mathbf{r})$ of component $\alpha$ at position $\mathbf{r}$, where $\alpha$ collectively denotes molecular species and their blocks. The total free energy functional reads \cite{uneyama_density_2005,uneyama_calculation_2005}
\begin{align}
      \frac{\mathcal{F}\left[\left\{\phi_{\alpha}(\mathbf{r})\right\}\right]}{\sqrt{\bar{\mathcal{N}}}k_{\rm B}T}
  &= \frac{1}{R_e^3}\int \mathrm{d}\mathbf{r} \left[P(\mathbf{r})N_0\left(\sum_{\alpha} \phi_{\alpha} - 1 \right) + \frac{1}{2} \sum_{\alpha \neq \beta} \chi_{\alpha\beta}N_0 \phi_{\alpha}(\mathbf{r}) \phi_{\beta}(\mathbf{r}) \right. \nonumber\\
  &\left.+\sum_{\alpha} A_{\alpha\beta} \sqrt{\phi_{\alpha}(\mathbf{r})} \int  \, \mathrm{d} \mathbf{r}^{\prime}\, \mathcal{G}(\boldsymbol{r}-\mathbf{r}^{\prime}) \sqrt{ \phi_{\beta}\left(\mathbf{r}^{\prime}\right)} \right. \nonumber  \\
  &\left.+\sum_{\alpha} C_{\alpha\beta} \phi_{\alpha}(\mathbf{r}) \ln \phi_{\alpha}(\mathbf{r})+\sum_{\alpha} \frac{R_e^2}{24}  \frac{\left|\nabla \phi_{\alpha}(\mathbf{r})\right|^{2}}{\phi_{\alpha}(\mathbf{r})} \right],
  \label{eq:continuum_hamiltonian} 
\end{align}
where $P(\mathbf{r})$ is a pressure-like Lagrange multiplier enforcing incompressibility, and $A_{\alpha\beta}$, $C_{\alpha\beta}$ are architecture-dependent coefficients. The second term describes enthalpic interactions via Flory-Huggins parameters $\chi_{\alpha\beta}$, the third term accounts for covalent bonding through a long-range interaction with kernel $\mathcal{G}(\mathbf{r})$, satisfying $(-\nabla^2+\xi_{\mathrm{cut}}^{-2})\mathcal{G}(\mathbf{r}) = \delta(\mathbf{r})$, the fourth term captures the entropy of mixing, and the fifth term imposes interfacial width via a gradient penalty.

For the AB diblock copolymer with block ratio $f_A = 1-f_B$, the architecture coefficients are
\begin{align}
    A = \frac{9N_0^2}{R_e^2f_A^2f_B^2}\left( \begin{matrix}
        f_B^2 & -f_A f_B \\ 
        -f_Af_B & f_A^2
    \end{matrix}\right), \quad
    C= \frac{N_0}{N_p} \left( \begin{matrix}
        \frac{\tilde s(f_A)}{f_B} & \frac{-1}{2\sqrt{f_Af_B}} \\ 
        \frac{-1}{2\sqrt{f_Af_B}} &  \frac{\tilde s(f_B)}{f_A}
    \end{matrix}\right),
\end{align}
with $\tilde s(f) = [s(f)-f]/[4f(1-f)]$. Here, the function $s(f) = 1.572 - 2.702 f(1-f)$ is a fitting function that ensures consistency with the order-disorder transition predicted by RPA \cite{ohta_equilibrium_1986}. For solvents and homopolymers, $A_\alpha = 0$ and $C_\alpha = N_0/N_\alpha$.

% The system comprises the same components as the particle simulations: AB diblock copolymer, volatile solvent S (THF), nonvolatile solvent C (DMF), nonsolvent N (water), and gas molecules G (air). The Flory-Huggins interaction matrix is identical to the particle model.

The temporal evolution is governed by model-B dynamics \cite{hohenberg_theory_1977} combining diffusive fluxes with conversion source/sink terms during the \ac{SNIPS} process. This results in \autoref{eq:model-B}, where $\mu_\alpha(\mathbf{r}) = \delta \mathcal{F}/\delta \phi_\alpha(\mathbf{r})$ is the chemical potential and $s_\alpha(\mathbf{r},t)$ describes molecular conversions. 

In the time evolution equation, the mobility, $\lambda_\alpha(\mathbf{r},t)/\lambda_0$, follows the same functional form as the inverse segmental friction in the particle model, $m_{A/B}$, accounts for vitrification. Weak thermal noise is incorporated through Gaussian random fields satisfying the fluctuation-dissipation relation, as described in Ref.~\cite{hafner_reaction-driven_2024}. In the here-presented simulations, thermal fluctuations have strength $\sqrt{\bar{\mathcal{N}}}=10^5$, only during the \ac{EISA} process. Hence, fluctuations are essentially not present, but weak lateral fluctuations are required in the first process step, for their amplitude to grow after the onset of microphase separation.
Conversions are performed both to model solvent evaporation, as well as the solvent-nonsolvent exchange. In both cases, these are performed in a conversion zone, which follows the interface position, $z_I$, at a constant distance of $\Delta z_{cz} = 0.6R_e$ during the \ac{EISA} process and is fixed during the \ac{NIPS} process. Hence the following conversion occurs exclusively for $z<z_I-\Delta z_{cz}$. For the \ac{EISA} process, solvent is converted to gas, $S\rightarrow G$, specifically $s_G = -s_S = r \phi_S$. For \ac{NIPS}, both the $S$ and $C$ solvent are exchanged for nonsolvent in the solution film, and therefore both are converted to nonsolvent at $z=0$. Here, $s_C = -r\phi_C$, $s_S = -r\phi_S$, and $s_N = r(\phi_S+\phi_C)$. In all cases the conversion occurs at a high rate, $r=10^3\lambda_0$, practically instantly converting all material. 

These continuum dynamics are flexibly and efficiently implemented within the \ac{UDM}-software in CUDA/C for GPU acceleration. The software is available open-source \textit{via} \url{www.gitlab.com/g.ibbeken/udm}.

\subsection{Particle-based Simulations}
\label{sec:particle_based_simulations-si}

The particle simulations employ a soft, coarse-grained model \cite{daoulas_single_2006} that represents several monomer repeat units by a single particle. Within the soft, coarse-grained model the Hamiltonian is split into strong bonded (b) and weak non-bonded (nb) interactions,
$
    \mathcal{H} = \mathcal{H}_\mathrm{b} + \mathcal{H}_{\mathrm{nb}}
$.

The strong bonded interactions are taken to be harmonic springs
\begin{align}
	\frac{\mathcal{H}_\mathrm{b}}{k_\mathrm{B}T} = \sum_m\sum_b \frac{3(N_0-1)}{2R_e^2} (\mathbf{r}_{m,b}-\mathbf{r}_{m,b+1})^2,
\end{align}
where $m$ indexes molecules and $b$ indexes bonds within each molecule. $\mathbf{r}_{m,b}$, $\mathbf{r}_{m,b+1}$ refer to the positions of the bonded particles. Further, $k_{\rm B}$ is the Boltzmann constant, $T$ the temperature, $N_0$ refers to the chain-contour discretization of the polymer, and $R_e$ to its root mean-squared end-to-end distance. 

The weak non-bonded interactions are expressed in terms of the normalized densities $\phi_\alpha(\mathbf{r})$ of component $\alpha$ at position $\mathbf{r}$, \textit{i.e.}, 
\begin{align}
	\frac{\mathcal{H}_{\mathrm{nb}}}{\sqrt{\bar{\mathcal{N}}}k_{\rm B}T} =
     \int \frac{\mathrm{d}\mathbf{r}}{R_e^3} 
    &\left(
    \frac{\kappa_0 N_0}{2} \left[\sum_{\alpha}
    \phi_\alpha(\mathbf{r}) - 1\right]^2 \nonumber \right. \\
    &\left.+ \frac{1}{2} \sum_{\alpha \neq \beta}
    \chi_{\alpha\beta}N_0 \phi_\alpha(\mathbf{r}) \phi_{\beta}(\mathbf{r}) 
    \right),
\end{align}
where $\kappa_0$ characterizes the inverse compressibility of the system, $\chi_{\alpha\beta}$ are the Flory-Huggins parameters describing the binary repulsion between the species $\alpha$ and $\beta$. The first term enforces near incompressibility and the second one accounts for the binary interactions of different components. $\sqrt{\bar{\mathcal{N}}} = nR_e^3/(VN_0)$ denotes the invariant degree of polymerization and sets the strength of thermal fluctuations. $n$ denotes the total number of beads in the cubic simulation cell of volume $V$ with periodic boundary conditions. The densities are calculated on a cubic collocation grid with linear spacing $\Delta x = \Delta y = \Delta z$.

The system comprises an AB diblock copolymer and multiple solvent types to model the \ac{SNIPS} process: a volatile solvent S (THF), a nonvolatile solvent C (DMF), a nonsolvent N (water), and gas molecules G (air). The interactions of the base parameter combination are given in the main text.

We employ the \ac{SCMF} algorithm \cite{daoulas_single_2006,schneider_multi-architecture_2019} that temporarily replaces the weak, non-bonded interactions by external fields and thereby exploits the different strengths of strong bonded and weak but computationally costly non-bonded interactions. Particle positions are updated by the smart-Monte-Carlo algorithm, using the strong bonded forces to bias the trial displacement. The time it takes a copolymer to diffuse its own mean end-to-end distance, $R_e$, in a disordered system serves as the time unit $\tau_R$.

To account for vitrification during the membrane formation process, polymer segments A and B are slowed down depending on the local concentration fields via the inverse segment friction $m(\{\phi_\alpha\})$. The inverse polymer segment friction varies with local concentration according to
\begin{align}
m(\{\phi_\alpha\}) = \frac{1}{2}\left[1 + \tanh\left(\frac{1 - \sum_\alpha a_\alpha \phi_\alpha}{3}\right)\right]
\label{eq:mobility}
\end{align}
with coefficients $a_\alpha$ the position, $\phi_P^*$ and width of the vitrification threshold, \textit{i.e.} at which polymer concentration the polymer dynamics become arrested. Specifically, $a_A=a_B=17$, $a_S=a_C=-40$, $a_G=a_N=0$ were chosen, resulting in $\phi_P^\ast\approx 0.72$. The solvent mobilities were fixed to the value $m=0.88$.

The \ac{SNIPS} process is modeled by solvent exchange. During \ac{EISA}, volatile solvent molecules S reaching a conversion zone at the film surface are converted to gas molecules G, resulting in film surface retraction. The conversion zone tracks the film surface position at a fixed offset. To initiate \ac{NIPS}, all gas molecules G are converted to nonsolvent N, and subsequently both solvents S and C are converted to N in the conversion zone. The positioning of the conversion zone occurs in the same manner as in the continuum model, described above.

We use the highly parallel and \ac{GPU}-accelerated software \ac{SOMA}\cite{schneider_multi-architecture_2019}, available open-source via \url{www.gitlab.com/InnocentBug/SOMA}.

\section{Gibbs triangle}
\label{sec:gibbs-triangle-testing-data}
\begin{figure}[h!]
    \centering
    \includegraphics[width=0.65\linewidth]{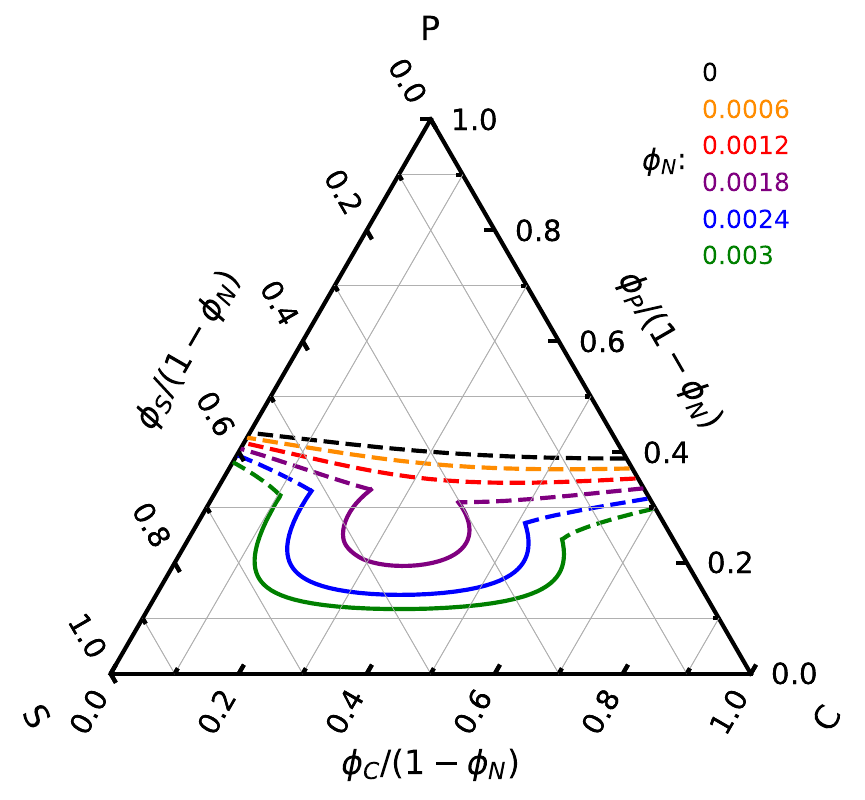}
    \caption{Spinodal curves of membrane casting solutions at different nonsolvent concentrations on the Gibbs triangle. The system parameters are identical to those used in \autoref{fig:error_plots}. Dashed and solid lines indicate instabilities of the spatially homogeneous state against composition fluctuations with finite and infinite length scales, respectively.}
    \label{fig:triangle}
\end{figure}

The Gibbs triangle corresponding to the random parameter combination of the testing data in \autoref{sec:error-prediction-results} is given in \autoref{fig:triangle}. From this and more resolved nonsolvent concentration $\phi_N$, the threshold concentration for macrophase separation to occur was determined to $\phi_N^\ast=0.0017$.

% \bibliographystyle{elsarticle-num}
% \bibliography{bibliography.bib}

\end{document}